\documentclass[prd,twocolumn,aps,reprint,nofootinbib]{revtex4-1}
\usepackage{amsmath,amssymb,amsfonts,graphicx,bm,color}
\usepackage[normalem]{ulem}
\usepackage[
colorlinks=true,
pdfproducer=medialab,
pdfa=true]{hyperref}
\usepackage{url}

\def\nn{\nonumber}
\def\l{\lambda}
\def\o{{\cal O}}
\def\br{{\cal B}}

\def\Bstomm{B_s \to \mu^+\mu^-}
\def\Btomm{B_d \to \mu^+\mu^-}

\def\rtm{\rho_{\tau\mu}}
\def\rmt{\rho_{\mu\tau}}

\def\rtt{\rho_{tt}}
\def\rbs{\rho_{bs}}
\def\rsb{\rho_{sb}}
\def\rbd{\rho_{bd}}
\def\rdb{\rho_{db}}
\def\rqb{\rho_{qb}}
\def\rbq{\rho_{bq}}

\def\cg{c_\gamma}

\begin{document}

\title{\boldmath Enhanced $B_q \to \ell\ell^\prime$
and $B \to (K, \pi) \ell\ell^\prime$ in light of $(g-2)_\mu$}
\author{
Wei-Shu Hou, Girish Kumar and Sven Teunissen}
\affiliation{
Department of Physics, National Taiwan University, Taipei 10617, Taiwan}

\bigskip
\begin{abstract} 
We study lepton flavor violating (LFV) $B$ decays in a 
general two Higgs doublet model with sub-TeV exotic scalars. 
Two different parameter spaces are explored: one dominated 
by extra top Yukawa coupling $\rho_{tt}$, the other by LFV 
couplings relevant for the muon $g-2$ anomaly. In the 
first case, flavor constraints such as $\ell \to 
\ell^\prime\gamma$, $h \to \ell\ell^\prime$ imply LFV $B$ 
decays are far below experimental sensitivities. The 
second case needs to be close to the alignment limit, 
but $B_q \to \tau\mu$ and $B \to (K, \pi) \tau\mu$ 
rates can lie within the sensitivities of Belle~II 
and LHCb Upgrade~II.
Neutral $B$ meson mixings and $B,\, K,\, \pi \to \ell \nu$ 
decays provide important flavor constraints on parameter 
space. $B$ decays involving $e$--$\tau$ violation are 
constrained by $\mu\to e$ processes.

\end{abstract}

\maketitle


\section{Introduction}

Flavor changing neutral {couplings (FCNCs)} in the 
Standard Model (SM) occur only beyond tree-level and the
corresponding rates are small due to suppression from GIM 
mechanism \cite{Glashow:1970gm} and vanishing neutrino masses.
But new physics (NP) beyond SM could have interactions that 
allow for sizable FCNC processes. Therefore, the precise 
measurements of rare FCNC decays serve as powerful probes 
of physics beyond SM. In this context, rare $B$ decays offer excellent opportunities as,
in addition to loop factors,
these are suppressed further by small CKM factors, 
while the $b$ quark mass is sufficiently large so long- 
and short-distance effects can be separated and extracted 
with reasonable precision.

Rare leptonic decays of $B$ mesons are advantageous 
for study, as all hadronic effects are contained in 
the decay constant, calculable by lattice 
QCD~\cite{Aoki:2021kgd}, hence the decay rates can be 
predicted with great precision. On the experimental 
front, there has been excellent progress with 
ever increasing precision. One example is the 
helicity-suppressed rare $\Bstomm$ decay, which provides 
one of the most sensitive probes of scalar NP interactions.

Based on full Run 1 and Run 2 data with 9~${\rm fb}^{-1}$ 
luminosity, LHCb~\cite{LHCb:2021vsc, LHCb:2021awg} 
reported the branching ratio,
\begin{align}
	{\cal B}(B_s 
\to \mu\mu) = \left(3.09^{+0.46\,+ 0.15}_{-0.43\, -0.11}\right) 
\times 10^{-9}, \quad [{\rm LHCb}]
\label{eq: LHCb-Bsmumu}
\end{align}
Subsequently, CMS~\cite{CMS:2022mgd} announced 
their Run 2 result, 
\begin{align}
 {\cal B}(B_s \to \mu\mu) = \left(3.83^{+0.38+ 0.19+ 0.14}
                _{-0.36-0.16 -0.13}\right)\times10^{-9},
    \,[{\rm CMS}]
\label{eq: CMS-Bsmumu}
\end{align}
based on 2016-2018 data with 140~${\rm fb}^{-1}$ 
integrated luminosity.
The central values of LHCb and CMS results differ by 
$1.2\sigma$, and lie on opposite sides of SM expectation,
$\br(\Bstomm)_{\rm SM} = (3.66 \pm 0.14) \times 10^{-9}$
\cite{Bobeth:2013uxa,Beneke:2019slt}; but within errors, 
both measurements agree with SM, thereby provide strong 
bounds on NP interactions. The CKM suppressed $\Btomm$ is 
not measured yet. The current $95\%$ C.L. upper limit 
$\br(\Btomm) < 1.9 \times 10^{-10}$~\cite{CMS:2022mgd} 
is still above the SM prediction at 
$\br(\Btomm)_{\rm SM} = (1.03 \pm 0.05) \times 
10^{-10}$~\cite{Beneke:2019slt}.

A more exquisite probe to hunt for NP is provided by lepton 
flavor violating (LFV) decays of $B$ mesons. Since LFV 
phenomena is practically absent in SM, any experimental 
detection will be unambiguous signals for NP. BaBar, Belle,
and more recently LHCb have searched for 
LFV $B$ decays; no evidence so far has been observed, 
which provide stringent limits.
In Table~\ref{tab: LFV-B-decays} we list current bounds, 
together with projected sensitivities in the near future 
on promising LFV $B$ decays.\footnote{We note in passing
that in certain NP scenarios
(e.g., those considered in
Refs.~\cite{Glashow:2014iga,Calibbi:2015kma,Feruglio:2016gvd})
lepton flavor nonuniversality can also lead to
lepton flavor violation
in $B$ decays. However, LHCb recently
reported \cite{LHCb:2022qnv,LHCb:2022zom} measurements
concerning the latter, 
finding no evidence of lepton universality breaking
in $b \to s \ell^+\ell^-$ decays.}

{
\begin{table*}[t]
\begin{center}
\begin{tabular}{l@{\hspace{4em}}l@{\hspace{4em}}l}
\hline
\hline
Decay mode  &  90\% C.L. Upper Limit  &  Future sensitivity   \\	
\hline
$B_s \to \tau^\pm\mu^\mp$	&  
$3.4\times 10^{-5}$ (LHCb \cite{LHCb:2019ujz}) 	&  
\hskip 1cm ...	\\
$B_d \to \tau^\pm\mu^\mp$	&  
$1.2\times 10^{-5}$ (LHCb \cite{LHCb:2019ujz}) 	&  
$3 \times 10^{-6}$ (LHCb~II~\cite{LHCb:2018roe})\\
$B^+ \to K^+\tau^+\mu^-$	 & 
$2.8\times 10^{-5}$ (BaBar \cite{BaBar:2012azg}) &  
$\sim 3 \times 10^{-6}$ (Belle~II \cite{Belle-II:2018jsg}) \\
$B^+ \to \pi^+\tau^+\mu^-$	&  
$4.5\times 10^{-5}$ (BaBar \cite{BaBar:2012azg}) 	&  
\hskip 1cm ...\\
\hline
$B_d \to  \tau^\pm e^\mp$ &
$1.6 \times 10^{-5}$ (Belle \cite{Belle:2021rod}) &
\hskip 1cm ... \\
$B^+ \to K^+ \tau^+ e^-$ &
$1.5\times 10^{-5}$ (BaBar \cite{BaBar:2012azg})  &
$\sim 2 \times 10^{-6}$ (Belle~II \cite{Belle-II:2018jsg})  \\
$B^+ \to \pi^+ \tau^+ e^-$ & 
$2.0\times 10^{-5}$ (BaBar \cite{BaBar:2012azg}) &
\hskip 1cm ... \\
\hline
$B_s \to \mu^\pm e^\mp$	&  
 $5.4 \times 10^{-9}$ (LHCb~\cite{LHCb:2017hag}) 	&  
 $3 \times 10^{-10}$ (LHCb~II~\cite{LHCb:2018roe})  	\\
$B_d \to \mu^\pm e^\mp$	&  
 $1.0 \times 10^{-9}$ (LHCb~\cite{LHCb:2017hag}) 	&  
 $9 \times 10^{-11}$ (LHCb~II~\cite{LHCb:2018roe})  	\\
$B^+ \to K^+ \mu^+ e^-$	&  
 $6.4\times 10^{-9}$ (LHCb~\cite{LHCb:2019bix}) 	&  
\hskip 1cm ...\\
$B^+ \to \pi^+ e^\pm \mu^\mp$ & 
$1.7 \times 10^{-7}$ (BaBar \cite{BaBar:2007xeb})&  \hskip 1cm ...
\\
\hline
\hline
\end{tabular}
\caption{
Summary of current experimental data on LFV $B$ decays 
considered in our analysis.}
\end{center}
\label{tab: LFV-B-decays}
\end{table*}
}

In this article, we explore the possible size of LFV $B$
decays associated with $b \to q \ell\ell^{\prime}\, 
(q=s, d)$ in one of the simplest extensions of SM, the 
general two Higgs doublet model (g2HDM)~\cite{Lee:1973iz}
(for a review of 2HDMs, see~\cite{Branco:2011iw}),
sometimes denoted as 2HDM Type III~\cite{Hou:1991un},
where the Lagrangian itself contains flavor changing neutral
couplings of exotic scalar bosons,  denoted as $\rho_{ij}$
(which are defined
in Eq.~\eqref{eq: Lag} later). In our 
study, we investigate two very different parameter space 
choices motivated by different phenomenological reasons.
In the first case, the NP Yukawa matrices are somewhat SM-like 
in strength; the largest coupling, just as in SM, is the 
top-related diagonal coupling $\rtt\lesssim \l_t$, where $\l_t
= \sqrt2 m_t/v$ is the SM top Yukawa coupling. Another assumption, 
supported by experiment~\cite{ParticleDataGroup:2020ssz}, is of 
small but finite mixing, denoted as $c_\gamma \,(\equiv \cos\gamma$),
between $CP$ even scalars in the model
(alignment). It was shown~\cite{Hou:2020itz} that 
lepton-related NP couplings $\rho_{\ell\ell'}$ are constrained 
to be {small}, due to bounds from $h \to \ell\ell^\prime$ 
decay and $\mu\to e$, $\tau\to \mu$ LFV processes. Driven 
by $\rho_{tt}$ (or $\rho_{tc}$), this scenario can realize 
electroweak baryogenesis (EWBG) for explaining the baryon 
asymmetry of the Universe(BAU)~\cite{Fuyuto:2017ewj, Fuyuto:2019svr},
providing strong motivation for 
experimental exploration.

In the second case, we adopt the alignment limit of $c_\gamma
 \to 0$, which then allows for sizable NP lepton Yukawa 
couplings related to $\mu$--$\tau$ sector. This scenario 
is usually invoked for NP explanation of the anomalous 
magnetic moment of muon, $(g-2)_\mu$, recently affirmed 
by Muon g-2 collaboration \cite{Muong-2:2021ojo}. In this 
scenario, as opposed to the first case, in order to satisfy  
bounds from LHC direct search for $gg \to \phi \to \tau\mu$ 
and the flavor bound of $\tau\to\mu\gamma$~\cite{Hou:2021sfl}, 
the $\rtt$ coupling cannot be substantial.
After evaluating constraints from neutral $B_q$ mixing 
and {$H^+$-induced} leptonic decays of $B, K, \pi$, we 
identify parameter space that can lead to significantly 
large rates of LFV $B$ decays, which are within reach of
upcoming measurements. Though we mostly focus on 
flavor violation in the $\mu$--$\tau$ sector, we will 
present g2HDM {expectations} for flavor violation 
in $\mu$--$e$ and $\tau$--$e$ sectors as well.

This article is organized as follows. In Sec.~\ref{sec: lag} 
we introduce the g2HDM Lagrangian and set up our notation. 
In Sec.~\ref{sec: case1}, we discuss Case~I; after discussing 
bounds on $\rtt$ from $B_q\to\mu\mu$ and $B_q$ mixing,
we present our results for LFV $B$ decays. In 
Sec.~\ref{sec: case2}, we discuss Case~II by first 
revisiting the one-loop solution to $(g-2)_\mu$ in g2HDM; 
after discussing the main constraints on relevant couplings,
we present our results for LFV B decays. Finally, 
in Sec.~\ref{sec: summary}, we present our conclusions.

\section{New Yukawa interactions}\label{sec: lag}
Adding a second Higgs doublet to the SM gives four 
new Higgs bosons, the neutral scalars $H$, $A$, 
and charged Higgs boson $H^\pm$ in mass basis. In the 
limit of $CP$ conserving scalar potential, the $H\,(A)$ 
boson is $CP$-even (odd). Due to absence of discrete $Z_2$ 
symmetry on Yukawa sector, it is not possible to diagonalize 
simultaneously the Yukawa matrices associated with the two 
Higgs doublets. As a result, the Yukawa Lagrangian of g2HDM 
contains Higgs FCNCs, giving rise to flavor violation at 
tree-level. Working in the so-called Higgs basis~\cite{Georgi:1978ri,Lavoura:1994fv,Botella:1994cs},
the Yukawa Lagrangian in g2HDM is given by~\cite{Davidson:2005cw,Mahmoudi:2009zx},
\begin{align}
\mathcal{L} = 
 - & \frac{1}{\sqrt{2}} \sum_{f = u, d, \ell} 
 \bar f_{i} \Big[\big(\lambda^f_i \delta_{ij} s_\gamma + \rho^f_{ij} c_\gamma\big) h \nn\\
 & + \big(\lambda^f_i \delta_{ij} c_\gamma - \rho^f_{ij} s_\gamma\big)H
    - i\,{\rm sgn}(Q_f) \rho^f_{ij} A\Big]  R\, f_{j} \nn\\
 & - \bar{u}_i\left[(V\rho^d)_{ij} R-(\rho^{u\dagger}V)_{ij} L\right]d_j H^+ \nn\\
 &- \bar{\nu}_i\rho^\ell_{ij} R \, \ell_j H^+
 +\text{H.c.},
\label{eq: Lag}
\end{align}
where indices $i, j$ denote the generation of fermion $f$,
$Q_f$ the corresponding electric charge, and $R (L) 
= (1\pm\gamma_5)/2$ are chiral projections. Note that NP 
Yukawa matrices $\rho^f$ are in general not Hermitian, hence
elements $\rho_{ij}$ can have arbitrary complex phases. 

The presence of tree-level Higgs FCNCs lead to potentially
dangerous flavor violating decays of SM Higgs boson, 
$h \to f_i f_j \,(i \ne j)$, which are severely bound 
by experiments. To evade such constraints, usually some 
$Z_2$ symmetry is imposed on NP Yukawa sector
to enforce {the natural flavor conservation 
condition~\cite{Glashow:1976nt}.} However, the vertex 
$h f_i f_j$ in g2HDM is proportional to the mixing angle 
$c_\gamma$, therefore with suppression due to sufficiently 
small $c_\gamma$, as hinted by current Higgs 
data~\cite{ParticleDataGroup:2020ssz}, the mere existence 
of Higgs FCNCs in g2HDM is not directly a cause of 
concern. But, of course, the strength of Higgs 
FCNC couplings will be determined by data.

The scalar potential can be found, e.g., in 
Ref.~\cite{Hou:2017hiw}. For our study, besides 
Eq.~\eqref{eq: Lag}, we only need physical $H, A,$ 
and $H^{+}$ masses as benchmarks. We focus on 
sub-TeV masses in range of $[300,\, 500]$ GeV and 
take $m_A=m_{H^+}$, usually adopted\footnote[2]{
Recent $M_W$ measurement by CDF~\cite{CDF:2022hxs} shows 
significant tension with SM, as well as measurements by 
other experiments. The CDF value can be explained in 
g2HDM (see, for example, Refs.~\cite{Bahl:2022xzi, 
Song:2022xts, Babu:2022pdn, Arco:2022jrt}) by inducing 
NP contribution to $T$ parameter. This, however, would 
necessarily require the masses of exotic scalars to be 
nondegenerate. Given the current situation is unclear, 
we do not consider accounting for the CDF result. 
}
to evade constraint from $T$ parameter (constraints from 
$S$ and $U$ parameters are easily satisfied~\cite{ONeil:2009fty,Davidson:2010xv}), where 
the formula for $T$ in g2HDM is given in appendix~\ref{app: T}.
In addition to oblique parameters, the parameter space 
considered in Sec.~\ref{sec: case1} and \ref{sec: case2} 
satisfy~\cite{Hou:2021sfl,Ghosh:2019exx} 
perturbativity, unitarity, and positivity. 

\section{Case I: top Yukawa dominance}\label{sec: case1}

We assume that NP top coupling $\rtt\sim \l_t$ is the 
largest coupling. This assumption finds support also from 
the {Cheng-Sher ansatz}~\cite{Cheng:1987rs}, 
$\rho_{ij} \propto \sqrt{m_im_j}/v$, frequently employed 
to control tree-level Higgs FCNC. But as mentioned in the 
preceding section, small $c_\gamma$ can tackle the issue 
of Higgs FCNC, so we do not quite follow the ansatz.
We take $c_\gamma\sim 0.1$ as sample value, but note that
due to $\rtt$ being the dominant quark coupling, the
main g2HDM contribution to $b \to q \ell\ell^{(\prime)}$
processes are induced by $H^+$ interactions that do 
not depend on $c_\gamma$. The $\cg$ value is relevant for 
constraints from $h \to \ell\ell^\prime$.

The leading flavor constraints on $\rho_{tt}$ are from 
B physics discussed later in detail.
For LFV $B$ decays, we also need to determine the strength of
lepton couplings $\rho_{\ell\ell^{\prime}}$.
We have discussed previously~\cite{Hou:2020itz}
the allowed strength of $\rho^\ell$ for large $\rtt$,
so let us give a brief summary. For finite $\cg$, 
$h\to \ell\ell^\prime$ provide important constraints 
on $\rho_{\ell\ell^\prime}$, independent of $\rtt$. For 
example, the current upper limit on $h \to \tau\mu$ 
from CMS, based on full Run 2 data~\cite{CMS:2021rsq},
\begin{align}
	{\cal B}(h \to \tau\mu) < 0.15 \%
	\quad {\rm (95\%\; C.L.)}
\label{eq: h2taumu-CMS}
\end{align}
implies $|\rtm c_\gamma| < 0.1 \l_\tau$ for $\rtm=\rmt$, 
giving $\rtm \sim \l_\tau$ for $\cg=0.1$ (see
appendix~\ref{app: hll'} for expressions in g2HDM). 
Even if $\cg\sim 0$ so the $h\to\tau\mu$ bound of 
Eq.~\eqref{eq: h2taumu-CMS} can be evaded, $\mu$--$\tau$ 
couplings together with sizable $\rtt$ unavoidably generate
$\tau\to\mu\gamma$ at two-loop via Barr-Zee 
diagrams~\cite{Barr:1990vd}. The recently updated bound of
$\tau\to\mu\gamma < 4.2 \times 10^{-8}$~\cite{Belle:2021ysv}
from Belle again gives $\rtm \sim \o(\l_\tau)$ for 
$\rtt\sim \l_t$ and $m_{H, A}\sim 300$ GeV. 

The current upper limits on the $\tau$--$e$ sector, e.g. 
$h\to \tau e$~\cite{CMS:2021rsq} and $\tau \to e 
\gamma$~\cite{BaBar:2009hkt} give relatively weak bounds, 
but couplings related to $e$--$\mu$ are strongly constrained 
by $\mu\to e\gamma$. The MEG bound $\mu\to e\gamma 
< 4.2 \times 10^{-13}$~\cite{MEG:2016leq} gives 
$\rho_{\mu e}\rtt \lesssim 0.4 \l_e\l_t$ for $m_{H, A} 
\sim 300$ GeV~\cite{Hou:2021qmf}. This again suggest that 
for {$\rtt\sim \l_t$}, strengths of $\mu$--$e$ flavor 
violating couplings are similar to SM electron Yukawa, $\l_e$. 
Concerning flavor conserving $\rho_{\ell\ell}$, measurements 
related to $h \to \mu\mu$~\cite{ATLAS:2020fzp,CMS:2020xwi} and 
$h\to \tau\tau$~\cite{CMS:2021sdq} imply that, for $\cg \sim 
0.1$, strengths of $\rho_{\mu\mu}$ and $\rho_{\tau\tau}$ are 
close to $\o(\l_\mu)$ and $\o(\l_\tau)$, 
respectively~\cite{Hou:2021zqq}. A very important insight 
concerning the strength of $\rho_{ee}$ came in 
Ref.~\cite{Fuyuto:2019svr}, where it was uncovered that, 
to evade constraints from electric dipole moment of electron 
measured by ACME~\cite{ACME:2013pal,ACME:2018yjb}, $\rtt$ and 
$\rho_{ee}$ should follow the pattern of $|\rho_{ee}/\rtt| 
\propto \l_e/\l_t$, again echoing SM-like strength for 
$\rho_{ee}$.

With discussion as delineated above, we take the following 
structure for NP lepton Yukawa matrix 
$\rho^\ell$~\cite{Hou:2020itz},
\begin{equation}
 \begin{aligned}
		\rho_{\ell\ell}\lesssim \o(\l_\ell); \ \rho_{e\ell}\lesssim \o(\l_e); \
	    \rho_{\tau\ell^\prime}\lesssim\o(\l_\tau)
	    \quad (\ell^\prime\ne e),
\end{aligned}
\label{eq: rho}
\end{equation}
which  will be our working assumption in estimating rates of 
LFV $B$ decays in Case I.

\begin{figure}[t]
\includegraphics[height=2.35cm,width=3.8cm]{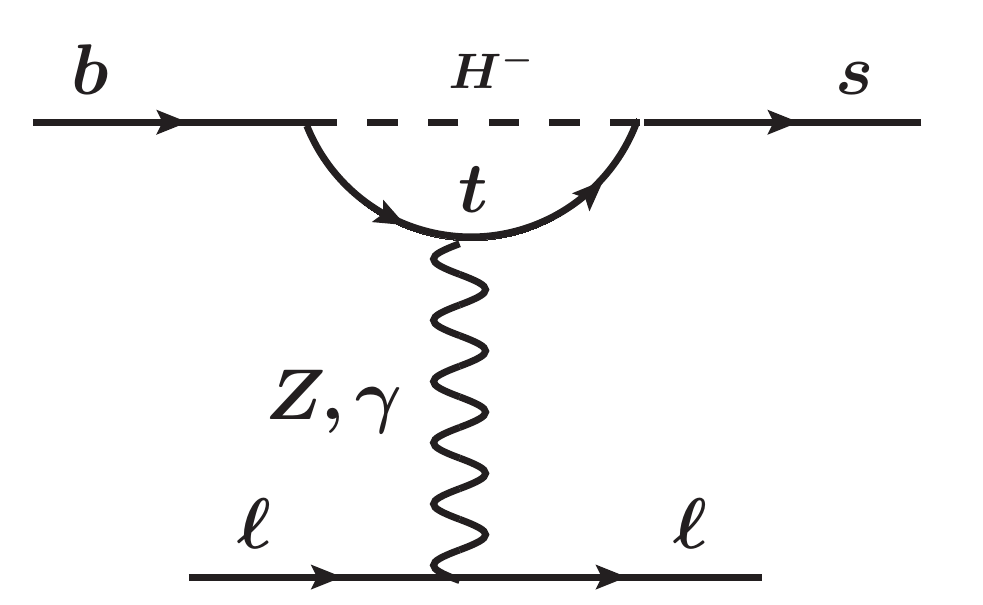}~~
\includegraphics[height=2.5cm,width=3.8cm]{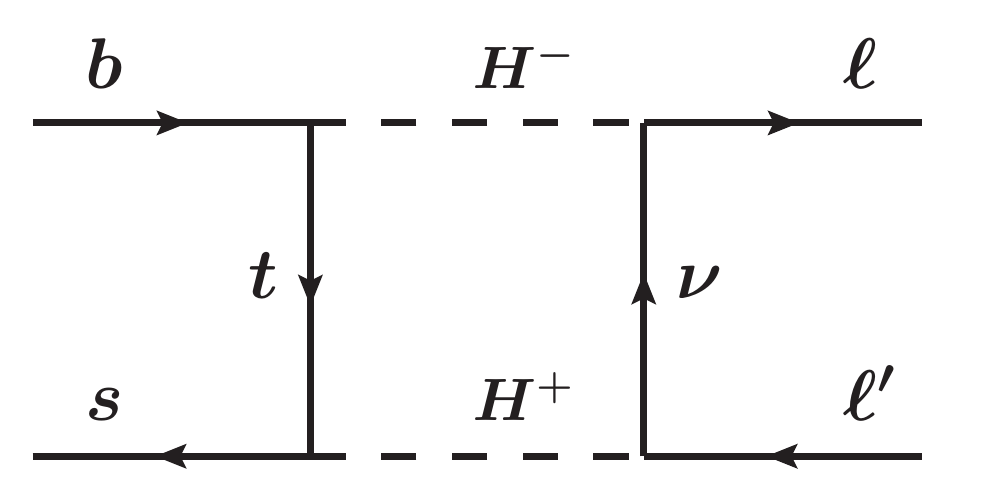}
\caption{$H^+$-induced Feynman diagrams for $b \to s\ell\ell$.}
\label{feyn: b2sll-loop}
\end{figure}

The NP contribution to $b\to q\ell\ell^{(\prime)}$
due to $\rtt$ arise from one-loop diagrams shown in
Fig.~\ref{feyn: b2sll-loop}, where the $Z$-penguin diagram
dominates. The diagrams generate the following
effective Hamiltonian, 
\begin{equation}
  - {\cal H}_{\rm eff}
	= C_V [\bar{q} \gamma_{\mu} L b]
	      [\bar{\ell} \gamma^{\mu} \ell^{(\prime)}]
	+ C_A [\bar{q} \gamma_{\mu} L b] 
	      [\bar{\ell} \gamma^{\mu}\gamma_{5} \ell^{(\prime)}],
\label{eq: Heff-1loop}
\end{equation}
contributing to both axial and vector coefficients,
\begin{align}
C^{Z}_{A} = \frac{V_{tq}^\ast V_{tb}|\rtt|^2}{16\pi^2 v^2} G_Z(x_t), 
\ \ 
	C^{Z}_{V} = -(1-4 s_W^2) C^{Z}_{A},
	\label{eq: b2sll-Z}
\end{align}
while $\gamma$-penguin contributes only to vector coefficient,
\begin{align}
	C^{\gamma}_{V} &= -\frac{e^2 V_{tq}^\ast V_{tb}|\rtt|^2}{16\pi^2 m_{H^+}^2}G_{\gamma}(x_t),
	\label{eq: b2sll-gam}
\end{align}
where $x_t=m_t^2/m_{H^+}^2$, $s_W$ is Weinberg angle and 
$G_{\gamma, Z}(x)$ are $t$-$H^+$ loop functions given in 
appendix~\ref{app: loop-fun}. Note that contributions in 
Eq.~\eqref{eq: b2sll-Z} and \eqref{eq: b2sll-gam} are 
universal to all lepton flavors due to SM vertex on lepton 
end, and therefore only affect lepton flavor conserving
$B$ decays.

The box diagram in Fig.~\ref{feyn: b2sll-loop}, however, 
does depend on lepton flavor and will contribute to 
LFV $B$ decays (box diagram with $W^+$-$H^+$ in the loop depends
on the down-type couplings and therefore does not contribute).
The corresponding Wilson coefficients are 
given by~\cite{Crivellin:2019dun},
\begin{align}
	C^{box}_{V} =C^{box}_{A}=
	\frac{-  V_{tq}^\ast V_{tb}|\rtt|^2 \rho^\ast_{i \ell}\,\rho_{i \ell^\prime}}
			{128\pi^2  m_{H^+}^2}[1+ 2\, G_Z(x_t)].
\end{align}
Before discussing numerical results, let us discuss briefly 
constraints on $\rtt$ from the flavor conserving decays 
$B_q\to\mu\mu$ and neutral $B_q$ mixings, which will help 
us determine the upper limit of $\rtt$ from data for a given 
value of $m_{H^+}$. For numerical analysis, we use the 
open-source packages \texttt{Flavio}~\cite{Straub:2018kue} 
and \texttt{Wilson}~\cite{Aebischer:2018bkb} for calculating 
flavor observables and QCD running of Wilson coefficients 
from NP scale to physical low-energy processes.

\begin{figure}[t]
\includegraphics[width=0.45\textwidth]{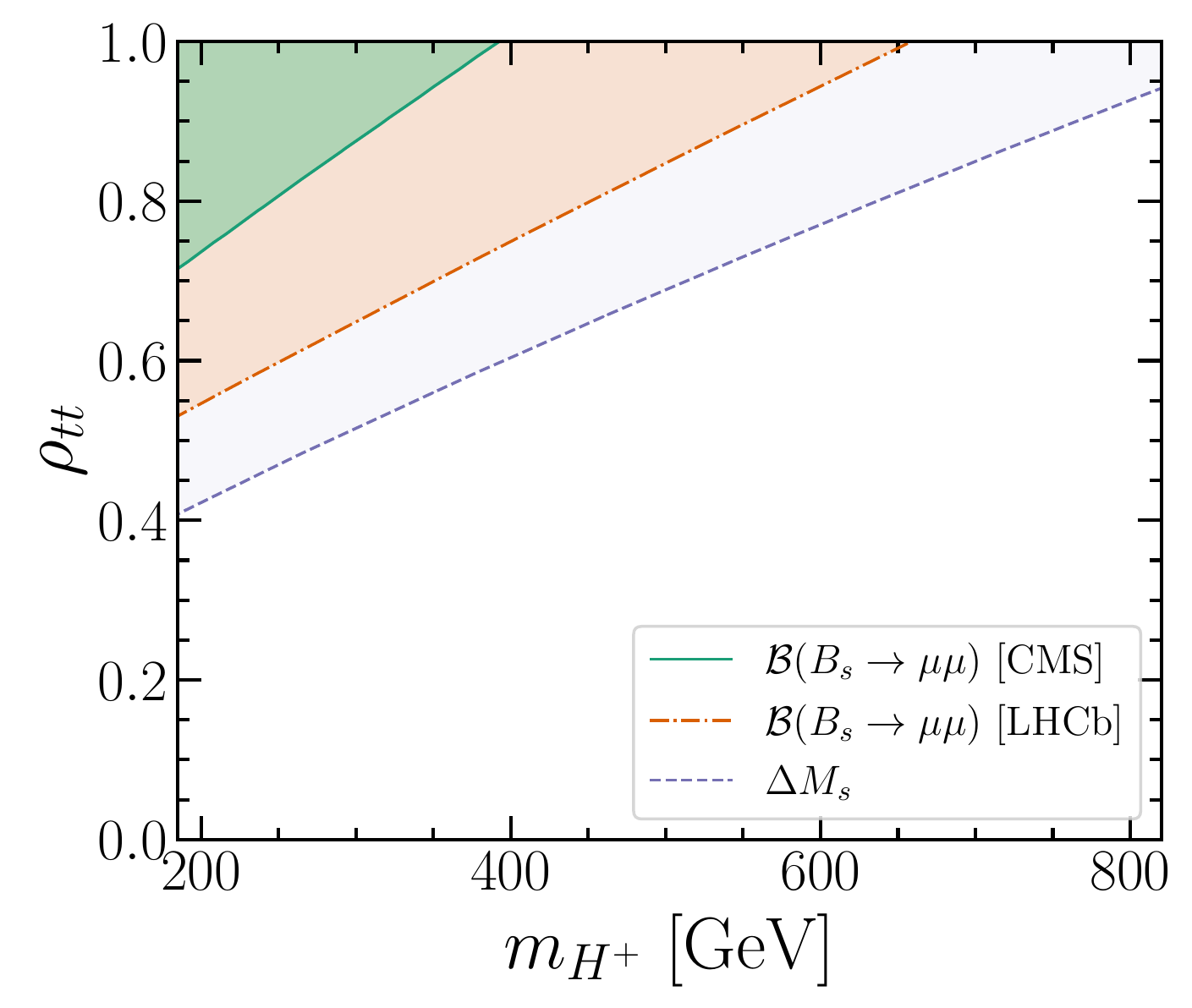}
\caption{Constraints in $\rtt$-$m_{H^+}$ plane from 
 B physics.}
\label{fig: B-physics-loop}
\end{figure}

In Fig.~\ref{fig: B-physics-loop}, we give the region ruled 
out at $95\%$ C.L. in $\rtt$--$m_{H^+}$ plane for both
LHCb~\cite{LHCb:2021vsc, LHCb:2021awg} (orange) and 
CMS~\cite{CMS:2022mgd} (green) measurements of $B_s \to 
\mu\mu$. It is interesting to note that, though the central 
value of the latest CMS measurement is closer to the SM 
prediction of $(3.66 \pm 0.14) \times 10^{-9}$, the resulting 
constraint on $\rtt$ is weaker compared to that from LHCb. 
This is because only the axial vector coefficient $C_A^Z$ 
modifies $B_s\to \mu\mu$ (see appendix~\ref{app: b2qll}) 
with the following correction,
\begin{equation}
 \frac{{\cal B}(B_{s}\to \ell \ell)}
      {{\cal B}(B_{s}\to \ell \ell)_{\rm SM}}
   \approx \left[1 - 1.2 \,|\rtt|^2 G_Z(x_t)\right]^2,
\label{eq: Bsmm-ratio}
\end{equation}
with $G_Z(x) < 0$. Since there is no sensitivity to 
$\arg\rtt$, $B_s\to \mu\mu$ rate can only be enhanced, 
so the central value of CMS being on the higher side
of SM allows for relaxed constraint on $\rtt$. Note 
also that, since Eq.~\eqref{eq: Bsmm-ratio} does not 
depend on CKM elements, the result holds true for $B_d 
\to \ell\ell$ as well. The current measurements for 
$B_q \to ee, \tau\tau$~\cite{ParticleDataGroup:2020ssz} 
are rather poor, therefore no improved constraint can be 
obtained. For $\rtt\sim 0.5$ and $m_{H^+}=300$ GeV, 
Eq.~\eqref{eq: Bsmm-ratio} implies that the rates of all 
$B_q\to \ell\ell$ get enhanced by $\sim 10\%$ over their 
SM value, which fits the {rising experimental trend}.

\begin{figure}[t]
\includegraphics[height=2.5cm,width=4cm]{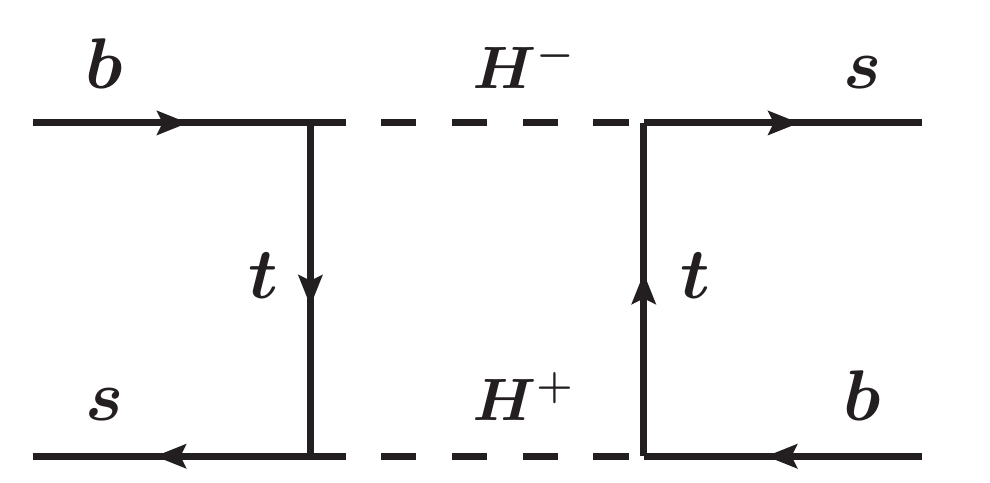}~~~
\includegraphics[height=2.5cm,width=4cm]{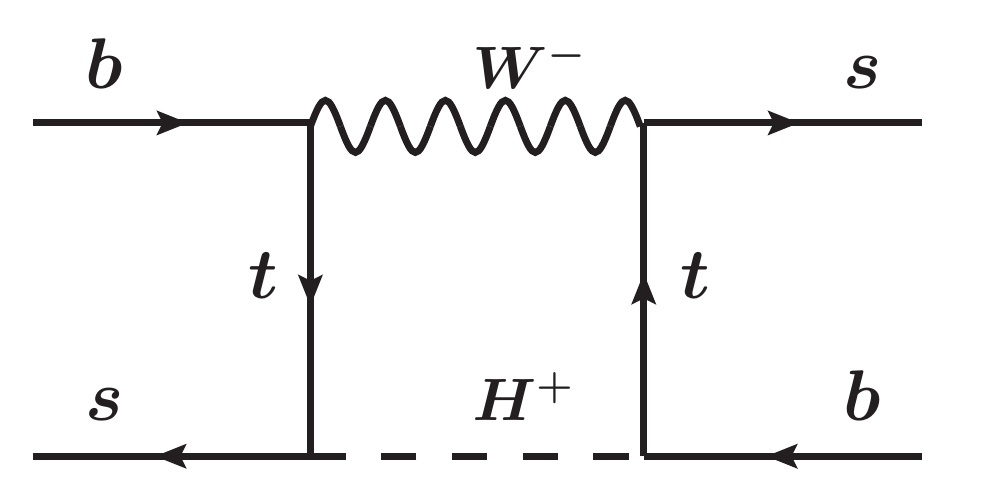}
\caption{$H^+$-induced Feynman diagrams for neutral B mixing.}
\label{feyn: Bmix-loop}
\end{figure}

A better constraint 
can be obtained from neutral $B_q$ mixings. The 
$\Delta B=\Delta S=2$ transitions arise from $H^+$ box 
diagrams of Fig.~\ref{feyn: Bmix-loop}, which 
generate~\cite{Crivellin:2019dun}, 
\begin{align}
 {\cal H}_{\rm eff} = (C_1^{HH} + C_1^{WH})
  [\bar s \gamma^\mu L b][\bar s \gamma^\mu L b] + {\rm h.c.}\, ,
\label{eq: Bmix-box-Heff}
\end{align}
where $C_1^{HH}$ is from $H^+$--$H^-$ diagrams,
\begin{align}
  C_1^{HH} = -\frac{V_{ts}^{\ast 2}V_{tb}^2 \,|\rho_{tt}|^4}
  			{128 \pi^2 m_{H^+}^2} f(x_t),
\label{eq: C1HH}
\end{align}
and $C_1^{WH}$ from $W^+$--$H^-$ diagrams,
\begin{align}
  C_1^{WH} = \frac{V_{ts}^{\ast 2} V_{tb}^2\,m_t^2|\rho_{tt}|^2}
  				  {32 \pi^2 v^2 m_W^2} g(y, x_t),
\label{eq: C1WH}
\end{align}
with $y = m_W^2/m_{H^+}^2$, and loop functions $f(x)$, $g(x)$ 
are given in appendix~\ref{app: loop-fun}. Similar expressions 
for $B^0$ mixing are obtained by replacing $s\to d$ in 
Eqs.~\eqref{eq: C1HH} and \eqref{eq: C1WH}.
The current values of mass differences $\Delta M_q$
are~\cite{ParticleDataGroup:2020ssz},
\begin{align}
\label{eq: DeltaMs}
	\Delta M_{B_s} &= (17.741 \pm 0.020)~{\rm ps}^{-1},\\
	\Delta M_{B_d} &= (0.5065 \pm 0.0019)~{\rm ps}^{-1},
	\label{eq: DeltaMd}
\end{align}
whereas SM predictions  are 
$\Delta M_{B_s} = (18.4^{+0.7}_{-1.2})~{\rm ps}^{-1}$
and $\Delta M_{B_d} = (0.533^{+0.022}_{-0.036})~{\rm ps}^{-1}$
\cite{DiLuzio:2019jyq}. 

In Fig.~\ref{fig: B-physics-loop}, the $\Delta M_s$ constraint
(light purple) is shown in $\rtt$--$m_{H^+}$ plane, which gives 
the leading constraint on $\rtt$. The constraints from $B^0$ 
mixing,
as well as from $b \to s \gamma$ (see appendix \ref{app: b2sgam}
for relevant NP contribution),
are relatively weak and not shown. Note also that, after 
replacing external fermion lines $\{bs\}\to \{sd\}$ in 
Fig.~\ref{feyn: Bmix-loop}, these box diagrams will contribute 
to neutral kaon mixing and modify mixing parameters $\Delta M_K$
and $\varepsilon_K$~\cite{ParticleDataGroup:2020ssz}, but the 
resulting constraints~\cite{Hou:2022qvx} on $\rtt$ are not 
competitive with $B_s$ mixing.

We find that $\rtt\sim 0.5$ is a reasonable choice 
with scalar mass spectrum in [300, 500]~GeV range. With 
lepton couplings from Eq.~\eqref{eq: rho}, we can now 
estimate various LFV B branching ratios. As discussed, both 
$Z$- and $\gamma$-penguins preserve lepton flavor, and only 
box diagrams of Fig.~\eqref{feyn: b2sll-loop} contribute. 
But this contribution is rather suppressed by small 
$\rho_{\ell\ell^\prime}$. For $\rtt=0.5$ and 
$m_{H^+}=300$ GeV, we find,
\begin{align}
	{\cal B}(B_s \to \mu\tau) \simeq  3 \times 10^{-18}, \quad 
	{\cal B}(B_d \to \tau\mu) \simeq  10^{-19},
\end{align}
with $B_q\to e \tau, e \mu$ further suppressed due to smaller
electron Yukawa couplings. We therefore find that LFV $B$ decays 
will be far below future sensitivities in 
Table~\ref{tab: LFV-B-decays} in g2HDM for Case I, with 
semileptonic decays $B \to (K, \pi) \ell\ell^\prime$ 
analogously suppressed.

\section{\boldmath Case II: $\mu$--$\tau$ Yukawa dominance
\label{sec: case2} }

The weakness of $\rho_{\ell\ell^\prime}$ in 
Eq.~\eqref{eq: rho}, together with GIM suppression, do not 
allow large LFV effects in Case I, but instead constrains 
$\rho_{tt}$ through $B_s \to \mu\mu$ and $B_s$ mixing. Now 
we explore a scenario where the LFV $\mu$--$\tau$ couplings 
can be sizable. One motivation for such parameter space is 
to address the disagreement between SM prediction and 
experimental measurement of the anomalous magnetic moment 
of the muon, $a_\mu = (g-2)_\mu/2$.

\begin{figure}[t]
\includegraphics[width=0.35\textwidth]{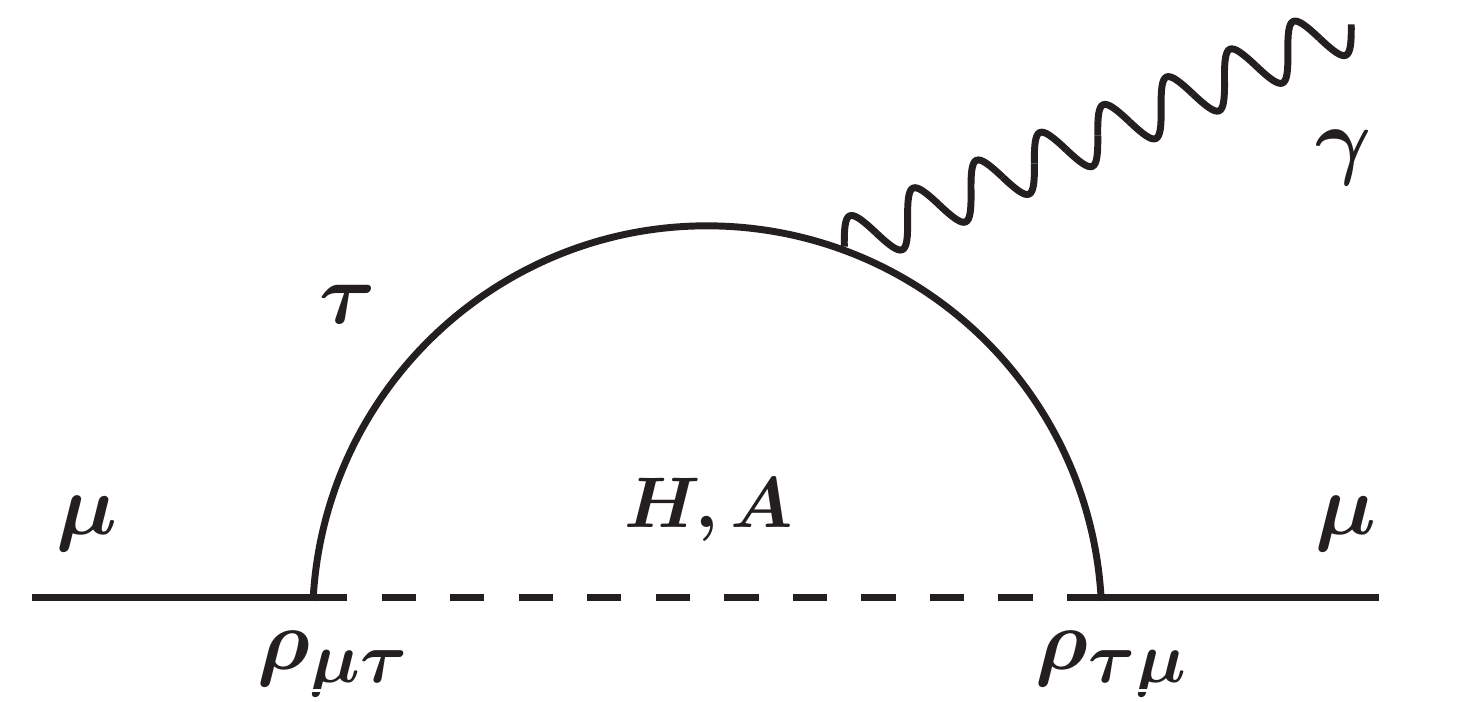}
\caption{One-loop diagram for $(g-2)_\mu$.}
\label{feyn: g-2}
\end{figure}

Recently, 
the Fermilab Muon g-2 experiment~\cite{Muong-2:2021ojo} 
reported its first measurement of $a_\mu$. Combined with 
the previous result of Brookhaven~\cite{Muong-2:2006rrc},
the result of $a_\mu^{\rm Exp} = 116 592 061(41) \times 10^{-11}$~\cite{Muong-2:2021ojo} compared to the theory 
consensus value of $a_\mu^{\rm SM} = 116 591 810(43) \times 
10^{-11}$~\cite{Aoyama:2020ynm, Aoyama:2012wk, Aoyama:2019ryr, Czarnecki:2002nt, Gnendiger:2013pva, Davier:2017zfy, 
Davier:2010nc, Keshavarzi:2018mgv, Colangelo:2018mtw, Hoferichter:2019mqg, Davier:2019can, Keshavarzi:2019abf, 
Kurz:2014wya, Melnikov:2003xd, Masjuan:2017tvw, 
Colangelo:2017fiz, Hoferichter:2018kwz, Gerardin:2019vio, 
Bijnens:2019ghy, Colangelo:2019uex, Blum:2019ugy,
Colangelo:2014qya} 
is larger by more
than $4\sigma$~\cite{Muong-2:2021ojo}\footnote[2]{The SM prediction
of $(g-2)_\mu$ based on the recent lattice results
\cite{Borsanyi:2020mff,Ce:2022kxy,Alexandrou:2022amy,Blum:2023qou,Bazavov:2023has}
is closer to the experimental value.
However, the low energy data on $\sigma (e^+e^-\to \mathrm{hadrons})$
\cite{Crivellin:2020zul,Keshavarzi:2020bfy,Colangelo:2020lcg} show
tension with these lattice results, which calls for further investigation.
In this paper we will take Eq.~\eqref{eq: Del-a_mu} as evidence of NP.
}
\begin{align}
\Delta a_\mu = a_\mu^{\rm Exp} - a_\mu^{\rm SM}
 =  (251 \pm 59) \times 10^{-11}.
\label{eq: Del-a_mu}
\end{align}
The difference can be explained in g2HDM via one-loop
diagram\footnote[3]{
In Case I, $\rtt$ together with $\rho_{\mu\mu}$ can 
contribute to $a_\mu$ at two-loop, but the contribution 
to $\Delta a_\mu$ is small~\cite{Hou:2021sfl}, due to 
constraint from $gg\to H/A \to \mu\mu$~\cite{CMS:2019mij, 
ATLAS:2019odt} direct search.
}
given in Fig.~\ref{feyn: g-2}, which in the limit of 
$c_\gamma\to 0$ gives the following NP 
correction~\cite{Assamagan:2002kf, Davidson:2010xv, 
Omura:2015xcg},
\begin{align}
	\Delta a _\mu|_\phi \simeq
	\frac{m_\mu m_\tau \operatorname{Re}(\rtm\rmt)}{16\pi^2 m_\phi^2 }
{\left[\log \frac{m_\phi^2}{m_\tau^2}
	          -\frac{3}{2}\right]},
	\label{eq: amu-1loop}
\end{align}
for each $\phi=H,\, A$. The total contribution is
$\Delta a _\mu = (\Delta a _\mu)_H - (\Delta a _\mu)_A$ as
H and A effects are opposite in sign. Therefore, to obtain
a finite $\Delta a_\mu$, $H$ and $A$ must be nondegenerate:
$\Delta m = m_A - m_H \ne 0$.

To present our numerical results, we follow 
Ref.~\cite{Hou:2021sfl} and assume $H$ to be lighter, 
setting $m_H = 300$~GeV. For $\Delta m$, we take two choices
for illustration: 40 and 200~GeV. The small $\Delta m = 
40$~GeV implies large cancellation between $H$ and $A$ 
contributions, and therefore a larger value of 
$\rtm\sim 30\l_\tau$ (we implicitly assume $\rtm=\rmt$) 
is required to account for difference in 
Eq.~\eqref{eq: Del-a_mu} within $1\sigma$ solution. For 
larger $\Delta m$ case, cancellation between $H$ and $A$ 
becomes mute since effect of heavy $A$ starts to decouple, 
and one only needs a smaller $\rtm=\rmt\sim 20 \l_\tau$.

With strength of $\rtm$ more than an order larger than 
Case~I (compare Eq.~\eqref{eq: rho}), the experimental 
bound of Eq.~\eqref{eq: h2taumu-CMS} implies $c_\gamma 
\sim 0.005$ or smaller. We therefore set $c_\gamma=0$ to 
simplify {(which would demand some yet unknown symmetry)}. 
Another important implication of large $\rtm$ is smallness 
of $\rtt$~\cite{Hou:2021sfl}, because of bound from 
$\tau\to\mu\gamma$~\cite{Belle:2021ysv}.
In fact, a more stringent constraint on $\rtt$ can 
be set~\cite{Hou:2021sfl} by the collider search for 
$gg \to H, A \to \tau\mu$~\cite{CMS:2019pex}. Therefore, 
if muon $g-2$ arises from one-loop in g2HDM, $\rtt$ is 
unavoidably small.

One possibility for enhancing LFV $B$ decays in g2HDM is 
to allow the $\rho^d$ Yukawa matrix to be nondiagonal. 
Explicitly, if one allows for finite $\rho_{bq}$ and 
$\rho_{qb}$ for $q=s,\,d$, $b\to q \ell\ell^{(\prime)}$ 
is at tree level while $B_q\to\ell\ell$ do not suffer 
helicity suppression. The effective Hamiltonian is,
\begin{equation}
	{\cal H}_{\rm eff} = - (C_S O_S + C_P O_P + C'_S O'_S + C'_P O'_P),
	\label{eq: Heff-tree}
\end{equation}
where
$\mathcal{O}_{S}  = (\bar s R b)(\bar\ell \ell'),$ 
$\mathcal{O}_{P}     = (\bar s R b)(\bar\ell \gamma_{5} \ell')$,
and ${\cal O}'_{S,P}$ are obtained by exchanging $L\leftrightarrow R$. 
The scalar Wilson coefficients at NP scale in alignment limit are given by,
\begin{align}
	C_{S,P}&=
	\frac{\rho_{sb}}{4}\left(\frac{\rho_{\ell\ell^\prime} \pm  \rho_{\ell^\prime\ell}^\ast}{m_H^2} - 
             \frac{\rho_{\ell\ell^\prime} \mp  \rho_{\ell^\prime\ell}^\ast}{m_A^2}\right),\\
             C_{S,P}^\prime  &=
	\frac{\rho_{bs}^\ast}{4}\left(\frac{\rho_{\ell\ell^\prime} \pm  \rho_{\ell^\prime\ell}^\ast}{m_H^2} + 
             \frac{\rho_{\ell\ell^\prime} \mp  \rho_{\ell^\prime\ell}^\ast}{m_A^2}\right).
 \label{eq:CSll'}
\end{align}
Note that under $\rho_{\ell\ell^\prime}=\rho_{\ell^\prime\ell}$ 
condition, {for each $C_{S, P}^{(\prime)}$ the $H$ and $A$ 
contributions are not simultaneously present.}

That down-type couplings $\rbq$, $\rqb$ can be finite and 
allowed from various flavor and collider constraints has 
been discussed~\cite{Crivellin:2017upt} for $h\to bq$ decays 
in g2HDM. The most important constraints on $\rho_{bq}$ 
couplings come from neutral $B_q$ mixings, which are now 
induced at tree-level. Ref.~\cite{Crivellin:2017upt} 
pointed out {an effective mechanism where if one 
imposes the conditions $\rho_{bq}\rho_{qb} = 0$ and $m_A 
= m_h m_H/\sqrt{m_h^2 s_\gamma^2 + m_H^2\cg^2}$, NP effects 
in $B_q$ mixing can be easily evaded.
}
However, note that the latter condition in the alignment 
limit implies $m_A=m_H$, which would rule out the possibility 
to explain the muon $g-2$ anomaly [Eq.~\eqref{eq: Del-a_mu}].
Therefore, one must confront $B_q$ mixing constraints in 
scenarios with $\Delta m \ne 0$.

\begin{figure}[b]
\includegraphics[width=.45\textwidth]{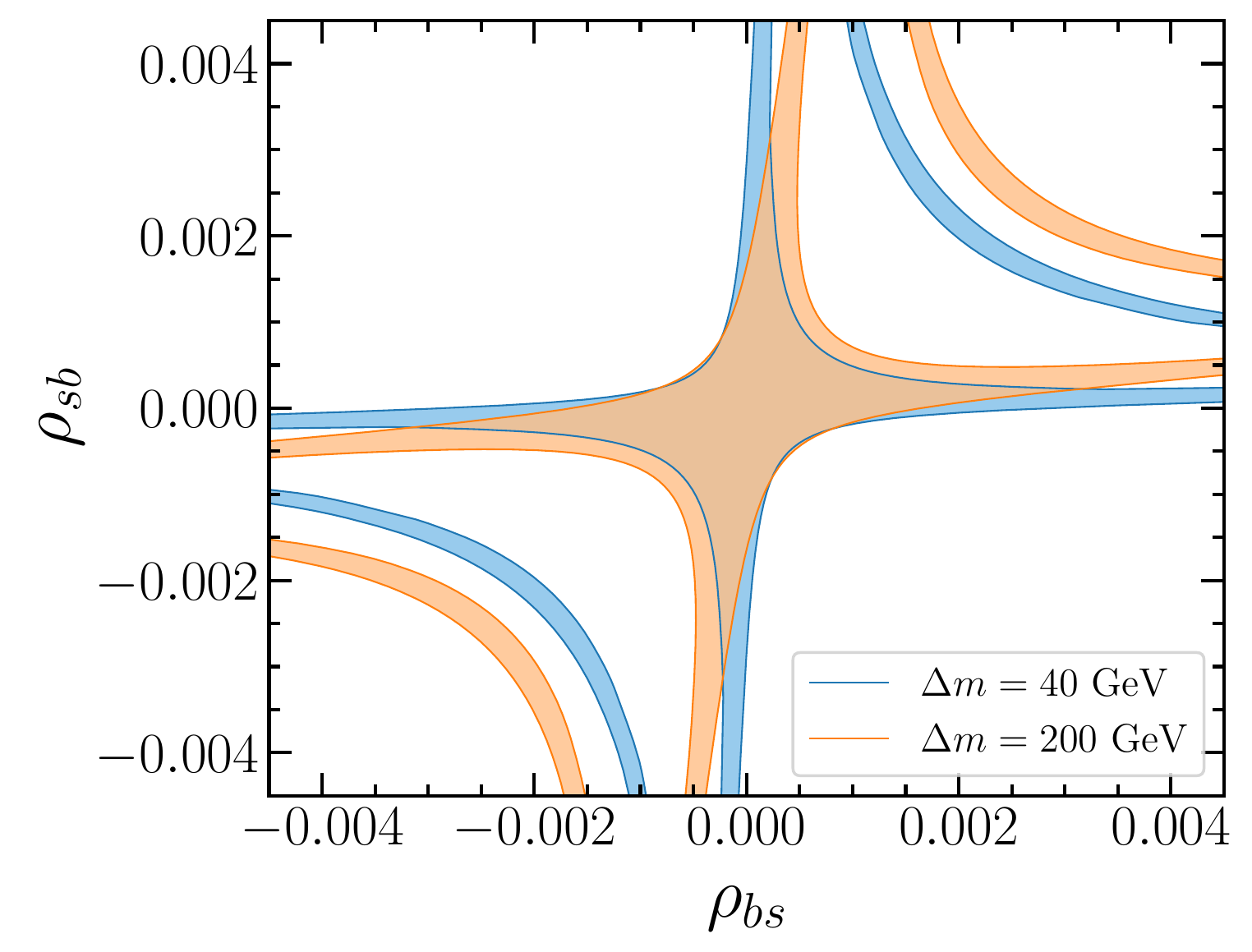}
\caption{Constraints from $B_s$ mixing.}
\label{fig: Bmix-tree}
\end{figure}

The couplings $\rbq$, $\rqb$ via tree-level $H/A$ exchange 
generate the effective Hamiltonian,
\begin{align}
 - {\cal H}_{\rm eff}
   = C_2 O_2 + C_2^\prime O_2^\prime + C_4 O_4,
\end{align}
where the Wilson coefficients at NP scale in alignment 
limit are given by,
\begin{align}\label{eq:B-mix-WC}
	C_2 = \frac{\rho_{bs}^{\ast 2}}{4}\left(\frac{1}{m_H^2}-\frac{1}{m_A^2}\right), ~
	C_4 = \frac{\rho_{bs}^\ast\rho_{sb}}{2}\left(\frac{1}{m_H^2}+\frac{1}{m_A^2}\right),
\end{align}
with $C_2^\prime$ obtained after substituting $\rbs^\ast 
\to \rsb$ in $C_2$.

In Fig.~\ref{fig: Bmix-tree}, we show the $95\%$ C.L. 
allowed region by $\Delta M_s$ measurement 
[Eq.~\eqref{eq: DeltaMs}] for $\Delta m= 40$ GeV (blue) 
and $200$ GeV (orange). One sees that, so long 
the product $\rbs\rsb$ is very small (but finite),
constraints on individual couplings $\rbs$ or $\rsb$ 
can be evaded (similar results follow for $\rbd$, $\rdb$ 
from $B_d$ mixing). Note, however, that if either $\rbs$ 
or $\rsb$ is zero, i.e. with $\rbs\rsb$ exactly zero, 
then the size of the other coupling is severely 
constrained and cannot be larger than $\o(10^{-3})$.

Another important probe for Case~II comes from $H^+$-induced 
processes. With lepton couplings fixed by 1$\sigma$ solution 
to muon $g-2$, $\rqb$ and $\rbq$ couplings contribute to 
leptonic decays such as $M^+ \to \ell^+\nu$ via tree-level 
$H^+$ exchange, where $M = B, K, \pi $, described by 
the effective Hamiltonian,
\begin{align} \label{eq: Heff-M2lnu}
	{\cal H}_{\rm eff}
	= -\frac{\rho^{\ast}_{\ell^{\prime}\ell}
	 \rho_{kj}^{d}V_{ik}}{m_{H^\pm}^2}
	\left(\bar{u}_i R d_j\right)\left(\bar{\ell} L \nu_{\ell^\prime}\right) + \text{H.c.},
\end{align}
which modifies the branching ratios as follows \cite{Hou:2019uxa},
\begin{align}
	&\frac{{\cal B}(M\to \ell\bar\nu)}{{\cal B}(M\to \ell\bar\nu)_{SM}}\nn\\
	& = \sum_{\ell^\prime}\left|\delta_{\ell\ell^\prime}
	- \frac{m_M^2 v^2\rho^{\ast}_{\ell^{\prime}\ell}\,\rho_{kj}^{d}
	V_{ik}}{2 V_{u_i d_j}(m_{u_i} + m_{d_j})\,m_\ell\, m_{H^\pm}^2 } \right|^2,
\label{eq: br-M2lnu}
\end{align}
where $m_M$ is the mass of meson $M$, and quark masses 
are evaluated at NP scale to account for renormalization 
group running. In Eq.~\eqref{eq: br-M2lnu}, neutrino 
species are summed over, since neutrino flavor is not 
detected by experiment {(the earlier work of 
Ref.~\cite{Crivellin:2013wna} contains an error here)}.

Decays $B\to \mu\nu$ and $B\to \tau\nu$ provide important
constraints on the coupling products $\rqb \rtm$ and 
$\rqb \rmt$, respectively. Adapting Eq.~\eqref{eq: br-M2lnu} 
for $B \to \mu \nu$, one notes that the SM-NP interference 
term (for $\ell^\prime = \mu$) involves the coupling 
$\rho_{\mu\mu}$, which is strongly constrained by 
$\tau \to \mu\mu\mu$~\cite{Hou:2021qmf}. Similarly, 
in case of $B \to \tau \nu$, the SM-interference term 
involves the coupling $\rho_{\tau\tau}$, which gets 
constrained by $\tau\to\mu\gamma$~\cite{Hou:2021qmf}. 
We therefore ignore the SM-NP interference term and 
focus on contributions of the coupling product $\rqb \rtm$ 
($\rqb \rmt$), which contribute through the incoherent 
term in ${\cal B} \to \mu\nu$ (${\cal B} \to \tau\nu$).

With current values of ${\cal B}(B\to\mu\bar\nu) 
= (5.3 \pm 2.0\pm 0.9)\times 10^{-7}$~\cite{Belle:2019iji}
and ${\cal B}(B\to\tau\bar\nu) = (1.09\pm 0.24) \times 
10^{-4}$~\cite{ParticleDataGroup:2020ssz}, we find the 
ratio $R_B^{\mu\tau} = \br(B\to \mu\nu)/\br(B\to \tau\nu)$ 
provides a better probe 
compared to individual branching ratios, as it is free 
from parametric uncertainties such as CKM elements and 
decay constant. In SM, one has $R_B^{\mu\tau}({\rm SM}) 
\simeq 0.0045$ with negligible errors, and using measured 
branching ratios, we obtain $R_B^{\mu\tau}({\rm exp}) 
= 0.0049 \pm 0.0023$. This value for $m_{H^+} = 340$~GeV 
gives $|\rsb\rtm| \lesssim 6.8 \times 10^{-4}$, and 
$|\rdb\rtm| \lesssim 1.55 \times 10^{-4}$. 
With $\rtm = \rmt = 0.3$ needed for $1\sigma$ solution 
to $\Delta a_\mu$, the coupling $\rho_{qb}$ is strongly 
constrained. But note that $\rbq$ remains unconstrained 
by $B\to \ell \nu$. Since $B_q$ mixing is ambivalent 
about which couplings, $\rqb$ or $\rbq$, is large, 
$B\to \ell\nu$ helps remove this ambiguity. That is, 
the coupling $\rbq$, compared to $\rqb$, is better 
suited for enhancing LFV $B$ decays.

\begin{figure*}[t]
\center
\includegraphics[width=0.48\textwidth]{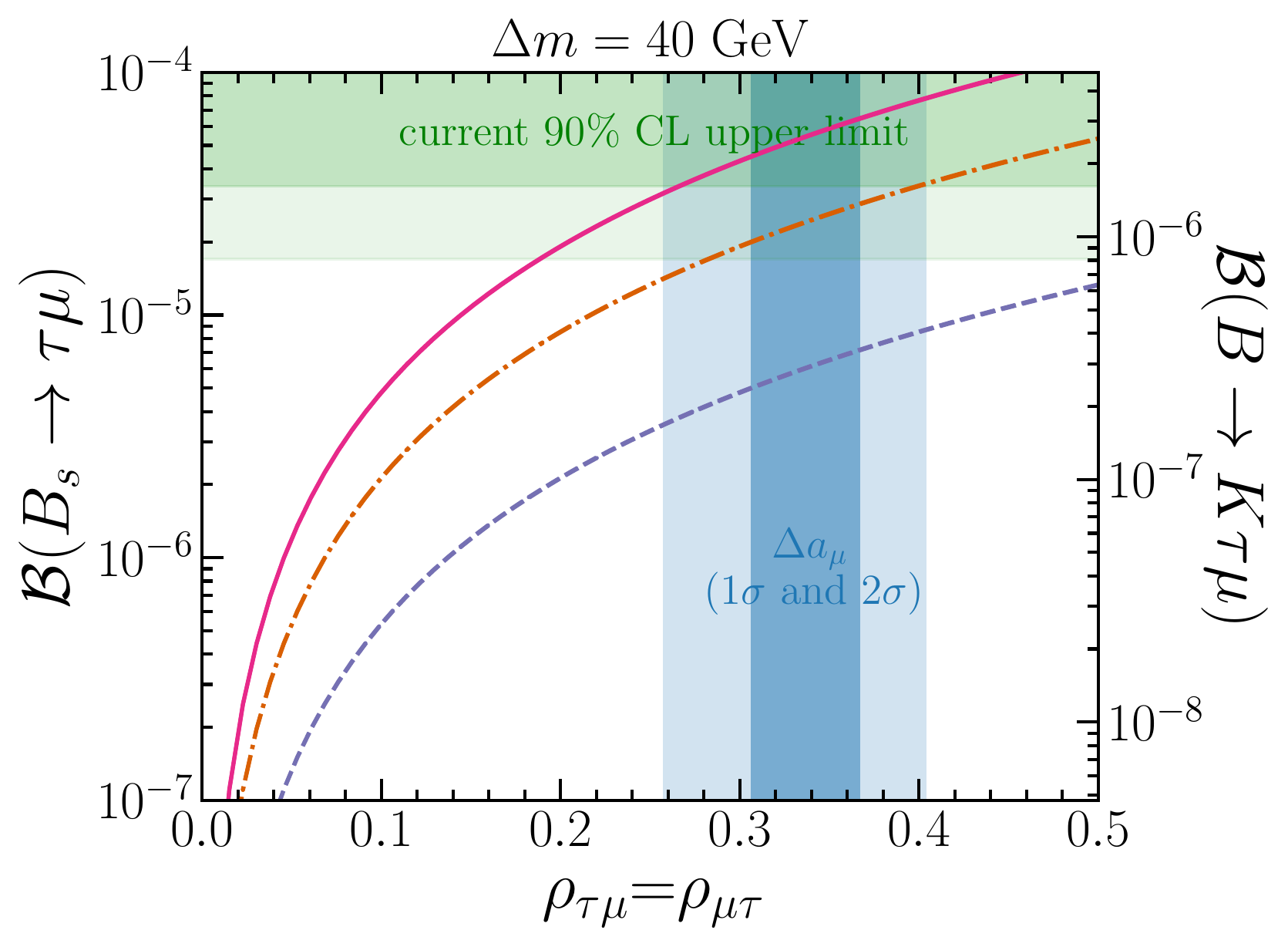}~~
\includegraphics[width=0.48\textwidth]{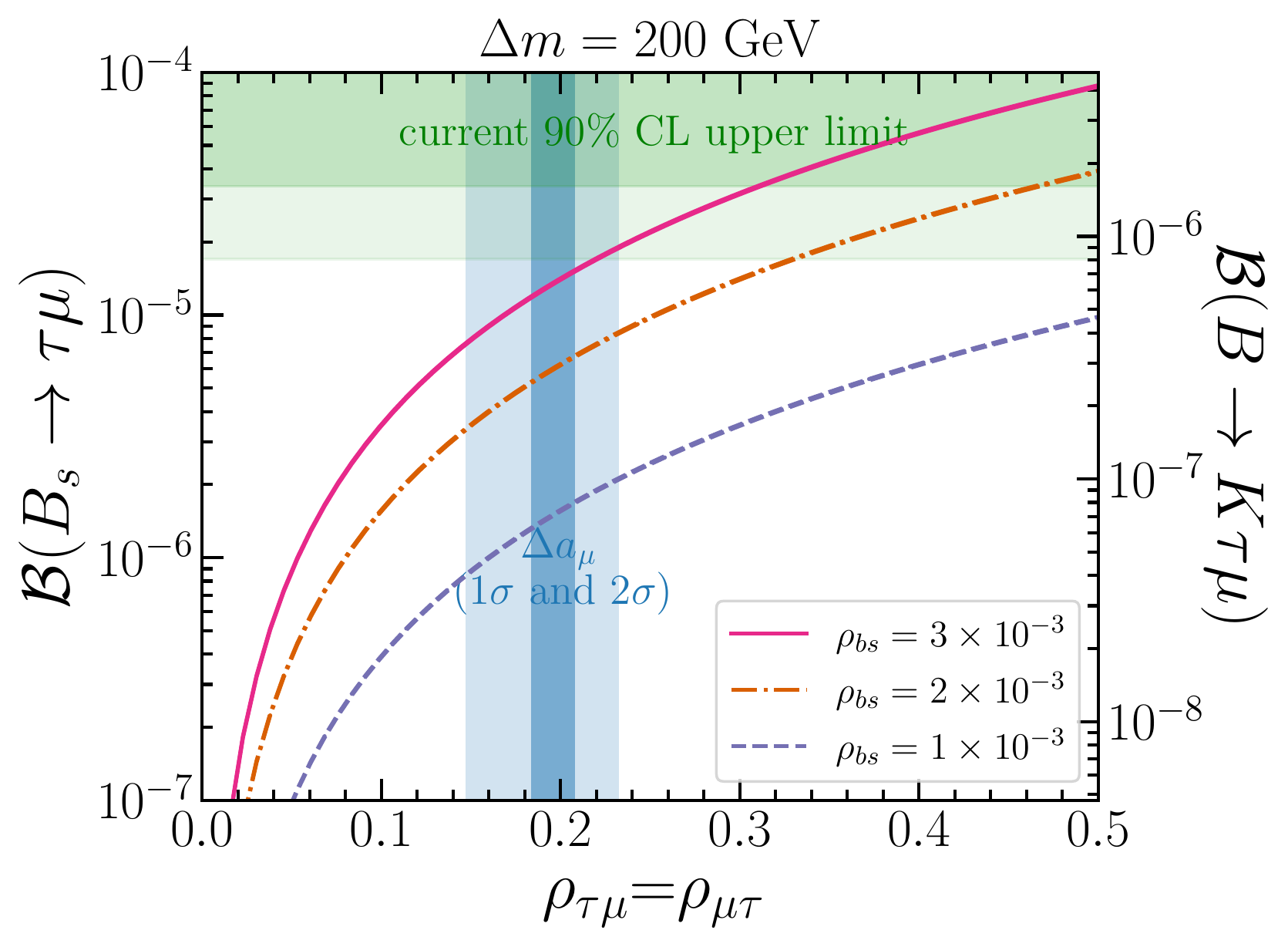}\\
\includegraphics[width=0.48\textwidth]{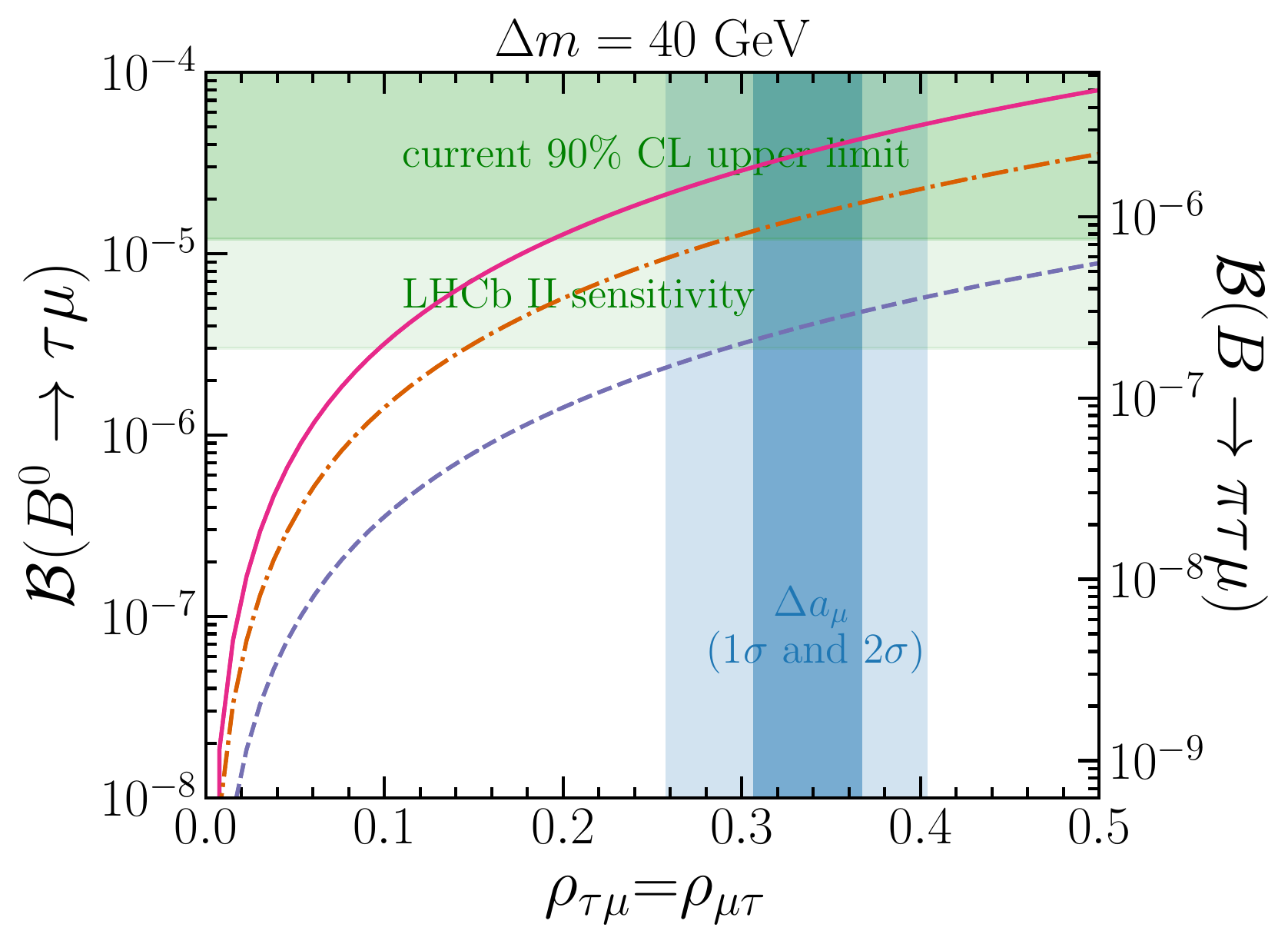}~~
\includegraphics[width=0.48\textwidth]{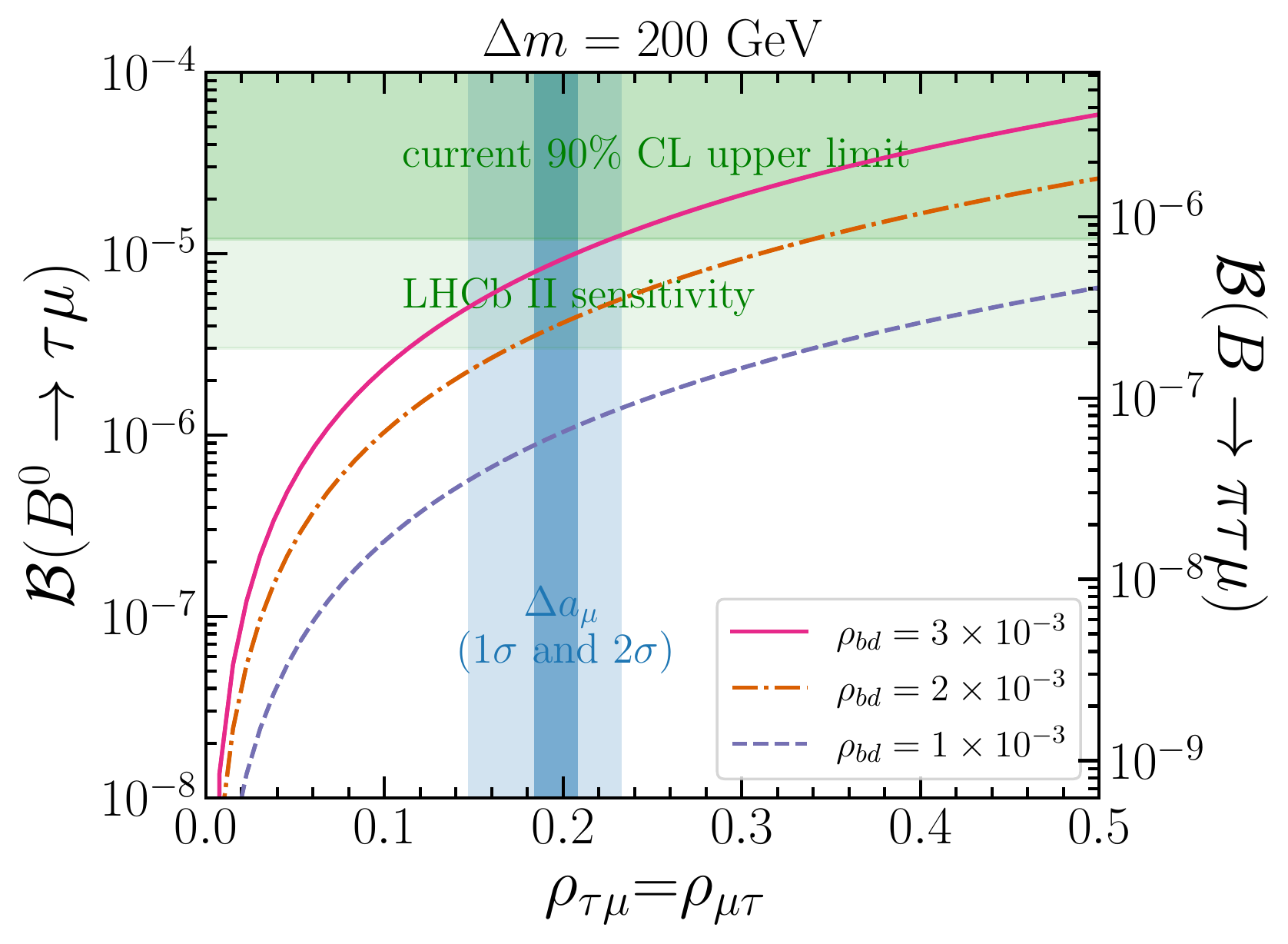}
\caption{Prediction for branching ratio of various LFV $B$ decays.}
\label{fig: br-LFV-B}
\end{figure*}

It is worth mentioning that the ratio $R_B^{\mu\tau}$ 
in 2HDM Type-II (such as in minimal supersymmetric models) 
is lepton flavor independent and therefore remains the
same as in SM. Therefore, the ratio $R_B^{\mu\tau}$ is 
one of the most important observables to probe genuine 
NP effects of g2HDM couplings~\cite{Hou:2019uxa}.
We mention in passing that constraints from other decays 
such as $K,\,D \to \mu\nu$ and $\tau \to (K, \pi)\nu$ do 
not impose any significant bounds.

Before presenting our results, we mention few important collider probes of Scenario II.
As noted in Ref.~\cite{Crivellin:2017upt}, if quark coupling $\rho_{bs}$
is large then pseudoscalar $A$ produced via strange-quark sea,
i.e., $s g \to b A$, followed by $A\to bs$ is one of the best channel to
search for.
However,  $\rho_{bs}\sim \mathcal{O}(10^{-3})$ is very small in our setup.
But lepton couplings $\rho_{\tau\mu}=\rho_{\mu\tau}$ are quite large.
Then exotic scalars $H, A$  can be probed with
 4-lepton final state (especially, the same-sign dimuon and same-sign ditau)
via electroweak scalar pair production:
$qq \to AH \to  \mu^\pm \mu^\pm \tau^\mp\tau^\mp$, as pointed out in
Ref.~\cite{Iguro:2019sly}. If the scalar pair is $H H^+, A H^+$ (or $H^+H^-$)
then 3-lepton plus neutrino (2-lepton plus 2 neutrinos) are also channels to
search for (see Ref.~\cite{Iguro:2019sly} for detail).
Another potential channel could be $bs \to H, A \to \mu\tau $.
Due to very small $\rho_{bs}$ the production cross-section
of $H, A$ at LHC is expected to be small, but given that strange quark
is involved, and that we also have sizable $\rho_{\tau\mu}=\rho_{\mu\tau}$,
it is not clear if this constraint can be ignored. We leave a detailed analysis
of collider signatures of scenario II as future work.

In Fig.~\ref{fig: br-LFV-B}, we present various LFV B 
decays as functions of $\rtm$ for a range of $\rbq$ values, 
while setting $\rqb=0$. The upper (lower) row shows results 
for $b\to s \tau\mu$ $(b \to d\tau\mu)$ related decays. 
Note that decays $B_q \to \ell\ell^\prime$ depend on the 
difference $C_i -C_i^\prime$, $i=S, P$ scalar Wilson 
coefficients, while semileptonic decays $B \to (K, \pi) 
\ell\ell^\prime$ depend on the sum $C_i + C_i^\prime$ 
{(see appendix \ref{app: b2qll})}. But since $C_{S, P}$ 
vanish because $\rqb=0$, only $C^\prime_{S, P}$ contribute. 
Thus, $B_q \to \ell\ell^\prime$ and $B 
\to (K, \pi) \ell\ell^\prime$ rates are correlated, 
as reflected in Fig.~\ref{fig: br-LFV-B}, where 
both modes are indicated for the $y$ axis. 
The dark green band in each plot corresponds to the region 
ruled out by current leptonic bounds. The light green 
region will be probed in the near future, according to 
Table~\ref{tab: LFV-B-decays}.\footnote[1]{
For $B_s\to \tau^\pm \mu^\mp$, due to lack of public 
results, we have conservatively assumed that future 
measurements can improve current limit at least
by a factor of 2.
}

We have not shown experimental sensitivity of
semileptonic decays as leptonic modes appear 
to be the leading probe.
The plots show that, for $\rbq\sim \o(10^{-3})$ 
and $\rtm\sim \o(20)\l_\tau$ as motivated by the 
muon $g-2$ anomaly, large rates of LFV $B$ decays are
possible and within reach of future searches.
{
One also notes from Fig.~\ref{fig: br-LFV-B} that 
scenarios with smaller mass splitting $\Delta m
< 200$~GeV have better prospects for discovery,}
{
although the needed $\rho_{\tau\mu}$ value is
larger, hence somewhat less attractive.}

Let us now briefly comment on $\tau$--$e$ and 
$\mu$--$e$ sectors. 
The $B$ decays with  $\tau$-$e$ flavor violation involve 
$\rho_{\tau e}$ and $\rho_{e\tau}$ couplings, but 
$\mu\to e \gamma$ puts a strong bound on them. 
The corresponding contribution to $\mu\to e\gamma$ is 
generated by a diagram similar to Fig.~\ref{feyn: g-2}, 
but with outgoing fermion replaced by electron. For values 
of $\rtm,\,\rmt$ that explain the $(g-2)_\mu$ anomaly,
the MEG bound of $\mu\to e\gamma < 4.2 \times 
10^{-13}$~\cite{MEG:2016leq} would imply $\rho_{\tau e} 
= \rho_{e\tau} \lesssim \o(\l_e)$~\cite{Hou:2021qmf}, 
which is quite severe. Therefore, to avoid the charged 
LFV constraint, we take $\rho_{\tau e}=\rho_{e\tau} 
\sim \l_e$. Then predictions with $\rbq=10^{-3}$, $m_H=300$ 
GeV, $m_A=340$ GeV are $\br(B_s \to \tau e) \sim 5 
\times 10^{-16}$, $\br(B \to K \tau e) \sim 10^{-17}$ 
and $\br(B_d \to \tau e) \sim 3 \times 10^{-16}$, 
$\br(B \to \pi \tau e) \sim 10^{-17}$, which are {far} 
below future sensitivities.

We find the coupling of the $\mu$-$e$ sector only weakly 
constrained in Case~II by charged LFV processes. The 
couplings $\rmt$, $\rtm$ together with $\rho_{\mu e}$, 
$\rho_{e\mu}$ contribute to $\tau\to e\gamma$ via diagrams 
similar to Fig.~\ref{feyn: g-2}, after replacing initial 
and final fermions by $\tau$ and $e$ and internal fermion 
by $\mu$. But the diagram is chirally suppressed by small 
$m_\mu$. Taking same mass as before and $\rho_{e\mu} = 
\rho_{\mu e}$, the current measurement of $\br(\tau \to 
e\gamma) = 3.3 \times 10^{-8}$~\cite{BaBar:2009hkt} sets 
the bound $\rtm\rho_{\mu e} \lesssim 
(5\times 10^5) \l_\tau\l_e$, which is quite poor. Note 
that $\tau^- \to \mu^-e^+\mu^-$ gives better constraint, 
as this decay is mediated by tree-level $H$, $A$ exchange 
hence does not suffer chiral suppression. Adapting the 
formula of $\tau^- \to \mu^-\mu^+\mu^- $ given in 
Ref.~\cite{Crivellin:2013wna} to $\tau^-\to \mu^-e^+\mu^-$, 
we find an order of magnitude improvement in constraint 
on $\rtm\rho_{\mu e}$ compared to $\tau\to e \gamma$.
Then taking $\rho_{e\mu}=\rho_{\mu e} \sim 10^3 \l_e \, 
(\simeq 0.003)$ and $\rbq=10^{-3}$ with $m_H=300$ GeV 
and $m_A = 340$ GeV, we find $\br({B_s \to \mu e}) \sim 
6 \times 10^{-10}$, $\br(B \to K \mu e) \sim 
3 \times 10^{-11}$  and 
$\br(B_d \to \mu e) \sim 4 \times 10^{-10}$, 
$\br(B \to \pi \mu e) \sim 3\times 10^{-11}$.
These values can be probed in the near future.

\section{Summary}\label{sec: summary}
We have explored prospects of enhanced lepton flavor 
violation in $B$ decays in g2HDM with sub-TeV exotic scalars. 
We focus on two different cases of parameter space. 
For Case~I, we assume the top Yukawa coupling $\rtt$ is 
the dominant quark coupling and take $c_\gamma\sim 0.1$. 
Then charged LFV processes $\tau\to\mu\gamma$ and 
$\mu\to e\gamma$ constrain $\rho_{\ell\ell^\prime}$ 
to Eq.~(5). Even with $\o(1)$ strength of $\rtt$, 
LFV $B$ decay rates are highly suppressed by small 
$\rho_{\ell\ell^\prime}$, and are far from the
sensitivities of upcoming LHCb Upgrade II and Belle II. 
For Case~II where $\mu$--$\tau$ flavor violating lepton 
couplings are motivated to be about 20-30 times larger 
than SM Yukawa $\l_\tau\simeq 0.01$. Contrary to Case~I,
one finds that $\rtt$ has to be small due to bounds from 
$\tau\to\mu\gamma$ and $gg\to H, A \to \tau\mu$ direct 
search at LHC. However, together with $\rbs,\,\rbd$ 
as small as $\o(10^{-3})$, we find that Case~II allows 
substantial rates of $\mu$--$\tau$ flavor violation in 
B decays, with B mixing and $B\to \mu\nu$ being the 
leading, but forgiving, flavor constraints. 

Concerning $\tau$--$e$ flavor violation in $B$ decays, 
we find that $\mu \to e \gamma$ would make it difficult 
to have simultaneously large $\rho_{\tau e}$, making 
$B_q\to \tau e$ and $B\to (K, \pi) \tau e$ rates too 
small to be probed at upcoming experiments. Furthermore, 
we find that current constraints on $\mu$--$e$ flavor 
violating coupling are not that severe. The future 
measurements of LFV B decays related to $b\to q \mu e$ 
will provide crucial constraint on g2HDM.

\vskip0.5cm
\noindent{\bf Acknowledgments}
This work is supported by 
NSTC 111-2639-M-002-002-ASP of Taiwan, 
and NTU grants 111L104019 and 111L894801.

\appendix
\section{\boldmath Useful Formulas}
\subsection{\boldmath The $T$-parameter}\label{app: T}
The parameter in g2HDM is defined as
\cite{ONeil:2009fty,Davidson:2010xv}
\begin{align}
T &= \frac{1}{16\pi s_W^2 m_W^2}\left\{
	   F(m_A^2, m_{H^+}^2)\right.\nn\\
  & \hskip1cm + \cg^2\left[F(m_{H^+}^2, m_h^2
                            - F(m_{A}^2, m_h^2)\right]\nn\\
  & \hskip1cm + s_\gamma^2
	\left[F(m_{H^+}^2, m_H^2)- F(m_{A}^2, m_H^2)\right]\nn\\
  & \hskip1cm - 3\, \cg^2
	 \left[F(m_{Z}^2, m_h^2)- F(m_{W}^2, m_h^2)\right. \nn\\
  & \hskip1cm + \left.\left.F(m_{W}^2, m_H^2)- F(m_{Z}^2, m_H^2)\right]\right\},
\end{align}
where function $F(a, b)$ is given by
\begin{align}
	F(a, b) = \frac{a+b}{2} - \frac{ab}{a-b}\log\frac{1}{b},
\end{align}
which vanishes in the limit $a\to b$.

\subsection{\boldmath $h \to \ell\ell^{(\prime)}$}
\label{app: hll'}

The tree-level decay rate for $h \to \ell\ell^\prime$ $(\ell\ne \ell^\prime)$ is,
\begin{align}
	\Gamma(h \to \ell\ell^\prime)
	&= \Gamma(h \to \ell^+\ell^{\prime -}) + \Gamma(h \to \ell^-\ell^{\prime +})\nn\\
	&\approx \frac{\cg^2 m_h}{16 \pi} (|\rho_{\ell\ell^\prime}|^2 + |\rho_{\ell^\prime\ell}|^2),
\end{align}
and for flavor conserving case,
\begin{align}
	\frac{\Gamma(h \to \ell\ell)}{\Gamma(h \to \ell\ell)_{\rm SM}} \approx \left|s_\gamma + \cg \,\operatorname{Re}\frac{\rho_{\ell\ell}}{\l_\ell}\right|^2 + \left|\cg \,\operatorname{Im}\frac{\rho_{\ell\ell}}{\l_\ell}\right|^2.
\end{align}

\subsection{\boldmath $B \to \ell\ell^{(\prime)}$,  $B \to M\ell\ell^{(\prime)}$ ($M = K, \pi$)}\label{app: b2qll}

The effective Hamiltonian for $b \to q\ell\ell^{(\prime)}$  is, 
\begin{equation}
	{\cal H}_{\rm eff} = -\frac{4G_F}{\sqrt{2}}V_{tq}^\ast V_{tb}\frac{e^2}{16\pi^2}\sum(C_i O_i + C_i^\prime O_i^\prime),
	\label{eq: Heff-b2qlll}
\end{equation}
with relevant operators,
\begin{align}
&O_{7} =\frac{m_b}{e} (\bar s \sigma^{\mu\nu} R b)F_{\mu\nu},~\, 
O_{8} =\frac{g_s^2}{e^2}{m_b} (\bar s \sigma^{\mu\nu} T^a R b)G^a_{\mu\nu},\nn \\
&O_{9} =\left(\bar{s} \gamma_{\mu} L b\right)
\left(\bar{\ell} \gamma^{\mu} \ell\right), \quad 
O_{10} =\left(\bar{s} \gamma_{\mu} L b\right)
\left(\bar{\ell} \gamma^{\mu}\gamma_5\ell\right),\nn \\
&O_{S} = \left(\bar{s} R b\right)
\left(\bar{\ell} \ell\right), \hspace{1cm}
O_{P} = \left(\bar{s} R b\right)
\left(\bar{\ell}\gamma_{5}\ell\right),
\label{eq: b2sll-ops}
\end{align}
where primed counterparts are obtained by $L\to R$ 
exchange. The full $b \to q \ell\ell$ operator basis can be found,
e.g., in Refs.~\cite{Grinstein:1988me, Buchalla:1995vs, 
Chetyrkin:1996vx}. One should take note of the
normalization used in ${\cal H}_{\rm eff}$ in 
Eq.~\eqref{eq: Heff-b2qlll} when comparing with 
${\cal H}_{\rm eff}$ defined in
Eq.~\eqref{eq: Heff-1loop} and
Eq.~\eqref{eq: Heff-tree} in the main text.
 
With NP operators of Eq.~\eqref{eq: b2sll-ops},
$\br(B_s\to \ell\ell^\prime)$ with respect to SM is
given by~\cite{Becirevic:2016zri}
\begin{align}
 &{\cal B}(B_{q}\to \ell \ell^{(\prime)}) =
     \frac{G_F^2 \alpha^2 |V_{tq}^\ast V_{tb}|^2
           f_{B_q}^2 \tau_{B_q}}{64\pi^3 m_{B_q}^3}
     \l^{\frac{1}{2}}(m_{B_q},m_{\ell},m_{\ell^\prime})\nn\\
     &\hskip0.1cm \times \Biggl\{(m_{B_q}^2 - m_+^2)
     \Bigg|\Delta C_9 \,m_{-}
     + \Delta C_S \frac{m_{B_q}^2}{m_b + m_q}\bigg|^2\Biggr.\nn\\
      & \hskip0.5cm \Biggl.
     + (m_{B_q}^2 - m_{-}^2)
     \bigg|\Delta C_{10}\, m_{+}
     + \Delta C_P \frac{m_{B_q}^2}{m_b + m_q}\bigg|^2\Biggr\},
\end{align}
where $\lambda(a, b, c) = [a^2-(b-c)^2][a^2-(b+c)^2]$,
and $m_\pm = m_\ell \pm m_{\ell^\prime}$, 
$\Delta C_i = C_i - C_i^\prime$.
For $\ell=\ell^\prime$, $C_9$ vanishes due to 
Ward identity for on-shell leptons.

The differential branching ratio of
$B\to (K, \pi) \ell\ell^{(\prime)}$ is
\begin{align}
 & d{\cal B}(B \to M\ell\ell^{(\prime)})/{dq^2}
    = |{\cal N}_M (q^2)|^2 \biggr\{\sum_{i}\varphi_i(q^2)
    |C_i + C_i^\prime|^2 \nn\\
 & \hskip1.2cm +\sum_{(i, j)}\varphi_{ij}(q^2) 
  \operatorname{Re}[(C_i + C_i^\prime)
   (C_j + C_j^\prime)^\ast]  \biggr\},
\end{align}
where $q$ is the $B$ to $M$ momentum transfer, and 
$i$ and $(i,j)$ run over $\{7, 9, 10, S, P\}$ and 
$\{(7, 9), (9, S), (10, P)\}$, respectively. The 
functions ${\cal N}_M(q^2)$ and $\varphi_{i(j)}(q^2)$ 
are given in Ref.~\cite{Becirevic:2016zri} (also see
Ref.\cite{Gratrex:2015hna} for a general formalism of
semileptonic $B$ decays).

\subsection{\boldmath $b \to s \gamma$}\label{app: b2sgam}
The $H^+$ induced dipole coefficients $C_7$ and $C_8$ mediating $b\to s \gamma$ and $b\to s g$ in g2HDM are given by, 
\begin{align}\label{eq: C78}
	\delta C_{7(8)}(x_t) =\frac{|\rho_{tt}|^2}{3|\lambda_t|^2} F_{7(8)}^{(1)}(x_t),
\end{align}
where the loop functions $F_{7 (8)}^{(1)}(x)$ are in the notation of
Ref.~\cite{Ciuchini:1997xe} (originally calculated in Ref.~\cite{Hou:1987kf})
and provided in the next appendix.

\subsection{\boldmath Loop functions}\label{app: loop-fun}
Loop functions related to $\Delta B=1$ decays \cite{Iguro:2017ysu}
and  $|\Delta B|=2$ processes are \cite{Crivellin:2019dun,Hou:2020itz}
are listed below.
\begin{widetext}
\begin{align}
G_Z(a) &= \frac{a(1-a+ \log a)}{2 (1-a)^2},\\
  G_{\gamma}(a) &= -\frac{2(16-45 a + 36 a^2 - 7 a^3 + 6(2-3a)\log a)}
  {108 (1-a)^4}\nn
  -\frac{2 -9a + 18 a^2 - 11 a^3 + 6a^3\log a}
  {36 (1-a)^4},\\
  F_7^{(1)}(a) &= \frac{a(7 - 5a - 8a^2)}{24(a-1)^3} + \frac{a^2(3a-2)}{4(a-1)^4}\log a,\\
	F_8^{(1)}(a) &= \frac{a(2 + 5a - a^2)}{8(a-1)^3} - \frac{3a^2}{4(a-1)^4}\log a,\\
	f(a) &=- \frac{1+a}{(a-1)^2} + \frac{2a\log{a}}{(a-1)^3}, \\
	g(a, b) &= \frac{1}{(a-b)^2}
	     \left[-\frac{3 a^2 \log{a}}{a-1}
                          + \frac{(b-4 a) (b-a)}{b-1}\right.
		            \;\ + \left.
		               \frac{\left(-4 a^2+3 a b^2 + 2ab - b^2\right) \log{b}}{(b-1)^2}\right].
\end{align}
\end{widetext}



\begin{thebibliography}{99}

\bibitem{Glashow:1970gm}
S.L.~Glashow, J.~Iliopoulos and L.~Maiani,
Phys. Rev. D \textbf{2}, 1285-1292 (1970).

\bibitem{Aoki:2021kgd}
Y.~Aoki, T.~Blum, G.~Colangelo, S.~Collins, M.~Della Morte, P.~Dimopoulos, S.~D\"urr, X.~Feng, H.~Fukaya and M.~Golterman, \textit{et al.}
[arXiv:2111.09849 [hep-lat]].

\bibitem{LHCb:2021vsc}
R.~Aaij \textit{et al.} [LHCb],
Phys. Rev. Lett. \textbf{128}, 041801 (2022)
[arXiv:2108.09284 [hep-ex]].

\bibitem{LHCb:2021awg}
R.~Aaij \textit{et al.} [LHCb],
Phys. Rev. D \textbf{105}, 012010 (2022)
[arXiv:2108.09283 [hep-ex]].

\bibitem{CMS:2022mgd}
 [CMS],
[arXiv:2212.10311 [hep-ex]].

\bibitem{Bobeth:2013uxa}
C.~Bobeth, M.~Gorbahn, T.~Hermann, M.~Misiak, E.~Stamou and M.~Steinhauser,
Phys. Rev. Lett. \textbf{112} (2014), 101801
[arXiv:1311.0903 [hep-ph]].

\bibitem{Beneke:2019slt}
M.~Beneke, C.~Bobeth and R.~Szafron,
JHEP \textbf{10} (2019), 232
[arXiv:1908.07011 [hep-ph]].


\bibitem{Glashow:2014iga}
S.L.~Glashow, D.~Guadagnoli and K.~Lane,
Phys. Rev. Lett. \textbf{114}, 091801 (2015)
[arXiv:1411.0565 [hep-ph]].

\bibitem{Calibbi:2015kma}
L.~Calibbi, A.~Crivellin and T.~Ota,
Phys. Rev. Lett. \textbf{115}, 181801 (2015)
[arXiv:1506.02661 [hep-ph]].

\bibitem{Feruglio:2016gvd}
F.~Feruglio, P.~Paradisi and A.~Pattori,
Phys. Rev. Lett. \textbf{118}, 011801 (2017)
[arXiv:1606.00524 [hep-ph]].



\bibitem{LHCb:2019ujz}
R.~Aaij \textit{et al.} [LHCb],
Phys. Rev. Lett. \textbf{123}, 211801 (2019)
[arXiv:1905.06614 [hep-ex]].

\bibitem{LHCb:2018roe}
R.~Aaij \textit{et al.} [LHCb],
[arXiv:1808.08865 [hep-ex]].

\bibitem{BaBar:2012azg}
J.P.~Lees \textit{et al.} [BaBar],
Phys. Rev. D \textbf{86}, 012004 (2012)
[arXiv:1204.2852 [hep-ex]].

\bibitem{Belle-II:2018jsg}
E.~Kou \textit{et al.} [Belle-II],
PTEP \textbf{2019}, 123C01 (2019)
[erratum: PTEP \textbf{2020}, 029201 (2020)]
[arXiv:1808.10567 [hep-ex]].



\bibitem{Belle:2021rod}
H.~Atmacan \textit{et al.} [Belle],
Phys. Rev. D \textbf{104}, L091105 (2021)
[arXiv:2108.11649 [hep-ex]].

\bibitem{LHCb:2017hag}
R.~Aaij \textit{et al.} [LHCb],
JHEP \textbf{03}, 078 (2018)
[arXiv:1710.04111 [hep-ex]].


\bibitem{LHCb:2019bix}
R.~Aaij \textit{et al.} [LHCb],
Phys. Rev. Lett. \textbf{123}, 241802 (2019)
[arXiv:1909.01010 [hep-ex]].

\bibitem{BaBar:2007xeb}
B.~Aubert \textit{et al.} [BaBar],
Phys. Rev. Lett. \textbf{99}, 051801 (2007)
[arXiv:hep-ex/0703018 [hep-ex]].
%

\bibitem{LHCb:2022qnv}
 [LHCb],
[arXiv:2212.09152 [hep-ex]].
%
\bibitem{LHCb:2022zom}
 [LHCb],
[arXiv:2212.09153 [hep-ex]].



\bibitem{Lee:1973iz}
T.D.~Lee,
Phys. Rev. D \textbf{8}, 1226 (1973).

\bibitem{Branco:2011iw}
G.C.~Branco \textit{et al.},
Phys. Rept. \textbf{516}, 1 (2012)
[arXiv:1106.0034 [hep-ph]].

\bibitem{Hou:1991un}
W.-S.~Hou,
Phys. Lett. B \textbf{296}, 179 (1992).

\bibitem{ParticleDataGroup:2020ssz}
P.A.~Zyla \textit{et al.} [Particle Data Group],
PTEP \textbf{2020}, 083C01 (2020).

\bibitem{Hou:2020itz}
W.-S.~Hou and G.~Kumar,
Phys. Rev. D \textbf{102}, 115017 (2020)
[arXiv:2008.08469 [hep-ph]].


\bibitem{Fuyuto:2017ewj}
K.~Fuyuto, W.-S.~Hou and E.~Senaha,
Phys. Lett. B \textbf{776}, 402 (2018)
[arXiv:1705.05034 [hep-ph]].

\bibitem{Fuyuto:2019svr}
K.~Fuyuto, W.-S.~Hou and E.~Senaha,
Phys. Rev. D \textbf{101}, 011901 (2020)
[arXiv:1910.12404 [hep-ph]].



\bibitem{Muong-2:2021ojo}
B.~Abi \textit{et al.} [Muon g-2],
Phys. Rev. Lett. \textbf{126}, 141801 (2021)
[arXiv:2104.03281 [hep-ex]].

\bibitem{Hou:2021sfl}
W.-S.~Hou, R.~Jain, C.~Kao, G.~Kumar and T.~Modak,
Phys. Rev. D \textbf{104}, 075036 (2021)
[arXiv:2105.11315 [hep-ph]].



\bibitem{Georgi:1978ri}
H.~Georgi and D.V.~Nanopoulos,
Phys. Lett. B \textbf{82}, 95 (1979).

\bibitem{Lavoura:1994fv}
L.~Lavoura and J.P.~Silva,
Phys. Rev. D \textbf{50}, 4619 (1994)
[arXiv:hep-ph/9404276 [hep-ph]].

\bibitem{Botella:1994cs}
F.J.~Botella and J.P.~Silva,
Phys. Rev. D \textbf{51}, 3870 (1995)
[arXiv:hep-ph/9411288 [hep-ph]].

\bibitem{Davidson:2005cw}
S.~Davidson and H.E.~Haber,
Phys. Rev. D \textbf{72}, 035004 (2005)
[erratum: Phys. Rev. D \textbf{72}, 099902 (2005)]
[arXiv:hep-ph/0504050].

\bibitem{Mahmoudi:2009zx}
F.~Mahmoudi and O.~Stal,
Phys. Rev. D \textbf{81}, 035016 (2010)
[arXiv:0907.1791 [hep-ph]].

\bibitem{Glashow:1976nt}
S.L.~Glashow and S.~Weinberg,
Phys. Rev. D \textbf{15}, 1958 (1977).

\bibitem{Hou:2017hiw}
W.-S.~Hou and M.~Kikuchi,
EPL \textbf{123}, 11001 (2018)
[arXiv:1706.07694 [hep-ph]].

\bibitem{CDF:2022hxs}
T.~Aaltonen \textit{et al.} [CDF],
Science \textbf{376}, no.6589, 170-176 (2022).

\bibitem{Bahl:2022xzi}
H.~Bahl, J.~Braathen and G.~Weiglein,
Phys. Lett. B \textbf{833}, 137295 (2022)
[arXiv:2204.05269 [hep-ph]].

\bibitem{Song:2022xts}
H.~Song, W.~Su and M.~Zhang,
[arXiv:2204.05085 [hep-ph]].

\bibitem{Babu:2022pdn}
K.S.~Babu, S.~Jana and V.P.~K.,
[arXiv:2204.05303 [hep-ph]].

\bibitem{Arco:2022jrt}
F.~Arco, S.~Heinemeyer and M.J.~Herrero,
[arXiv:2207.13501 [hep-ph]].

\bibitem{ONeil:2009fty}
D.~O'Neil,
[arXiv:0908.1363 [hep-ph]].

\bibitem{Davidson:2010xv}
S.~Davidson and G.J.~Grenier,
Phys. Rev. D \textbf{81}, 095016 (2010)
[arXiv:1001.0434 [hep-ph]].

\bibitem{Ghosh:2019exx}
D.K.~Ghosh, W.-S.~Hou and T.~Modak,
Phys. Rev. Lett. \textbf{125}, 221801 (2020)
[arXiv:1912.10613 [hep-ph]].

\bibitem{Cheng:1987rs}
T.-P.~Cheng and M.~Sher,
Phys. Rev. D \textbf{35}, 3484 (1987).



\bibitem{CMS:2021rsq}
A.M.~Sirunyan \textit{et al.} [CMS],
Phys. Rev. D \textbf{104}, 032013 (2021)
[arXiv:2105.03007 [hep-ex]].

\bibitem{Barr:1990vd}
S.M.~Barr and A.~Zee,
Phys. Rev. Lett. \textbf{65}, 21 (1990)
[erratum: Phys. Rev. Lett. \textbf{65}, 2920 (1990)].

\bibitem{Belle:2021ysv}
A.~Abdesselam \textit{et al.} [Belle],
JHEP \textbf{10}, 19 (2021)
[arXiv:2103.12994 [hep-ex]].


\bibitem{BaBar:2009hkt}
B.~Aubert \textit{et al.} [BaBar],
Phys. Rev. Lett. \textbf{104}, 021802 (2010)
[arXiv:0908.2381 [hep-ex]].

\bibitem{MEG:2016leq}
A.M.~Baldini \textit{et al.} [MEG],
Eur. Phys. J. C \textbf{76}, 434 (2016)
[arXiv:1605.05081 [hep-ex]].

\bibitem{Hou:2021qmf}
W.-S.~Hou and G.~Kumar,
Eur. Phys. J. C \textbf{81}, 1132 (2021)
[arXiv:2107.14114 [hep-ph]].

\bibitem{ATLAS:2020fzp}
G.~Aad \textit{et al.} [ATLAS],
Phys. Lett. B \textbf{812}, 135980 (2021)
[arXiv:2007.07830 [hep-ex]].

\bibitem{CMS:2020xwi}
A.M.~Sirunyan \textit{et al.} [CMS],
JHEP \textbf{01}, 148 (2021)
[arXiv:2009.04363 [hep-ex]].

%
\bibitem{CMS:2021sdq}
A.~Tumasyan \textit{et al.} [CMS],
JHEP \textbf{06}, 012 (2022)
[arXiv:2110.04836 [hep-ex]].

\bibitem{Hou:2021zqq}
W.-S.~Hou, G.~Kumar and S.~Teunissen,
JHEP \textbf{01}, 092 (2022)
[arXiv:2109.08936 [hep-ph]].


\bibitem{ACME:2013pal}
J.~Baron \textit{et al.} [ACME],
Science \textbf{343}, 269 (2014)
[arXiv:1310.7534 [physics.atom-ph]].

\bibitem{ACME:2018yjb}
V.~Andreev \textit{et al.} [ACME],
Nature \textbf{562}, 355 (2018).

\bibitem{Crivellin:2019dun}
A.~Crivellin, D.~M\"uller and C.~Wiegand,
JHEP \textbf{06} (2019), 119
[arXiv:1903.10440 [hep-ph]].


\bibitem{Straub:2018kue}
D.M.~Straub,
arXiv:1810.08132 [hep-ph].

\bibitem{Aebischer:2018bkb}
J.~Aebischer, J.~Kumar and D.M.~Straub,
Eur. Phys. J. C \textbf{78}, 1026 (2018)
[arXiv:1804.05033 [hep-ph]].


\bibitem{DiLuzio:2019jyq}
L.~Di Luzio, M.~Kirk, A.~Lenz and T.~Rauh,
JHEP \textbf{12}, 009 (2019)
[arXiv:1909.11087 [hep-ph]].

\bibitem{Hou:2022qvx}
W.-S.~Hou and G.~Kumar,
[arXiv:2207.07030 [hep-ph]].



\bibitem{Muong-2:2006rrc}
G.W.~Bennett \textit{et al.} [Muon g-2],
Phys. Rev. D \textbf{73}, 072003 (2006)
[arXiv:hep-ex/0602035 [hep-ex]].


\bibitem{Aoyama:2020ynm}
T.~Aoyama  \textit{et al.},
Phys. Rept. \textbf{887}, 1 (2020)
[arXiv:2006.04822 [hep-ph]].


\bibitem{Aoyama:2012wk}
T.~Aoyama, M.~Hayakawa, T.~Kinoshita and M.~Nio,
Phys. Rev. Lett. \textbf{109}, 111808 (2012)
[arXiv:1205.5370 [hep-ph]].

\bibitem{Aoyama:2019ryr}
T.~Aoyama, T.~Kinoshita and M.~Nio,
Atoms \textbf{7}, 28 (2019).

\bibitem{Czarnecki:2002nt}
A.~Czarnecki, W.J.~Marciano and A.~Vainshtein,
Phys. Rev. D \textbf{67}, 073006 (2003)
[erratum: Phys. Rev. D \textbf{73}, 119901 (2006)]
[arXiv:hep-ph/0212229 [hep-ph]].

\bibitem{Gnendiger:2013pva}
C.~Gnendiger, D.~St\"ockinger and H.~St\"ockinger-Kim,
Phys. Rev. D \textbf{88}, 053005 (2013)
[arXiv:1306.5546 [hep-ph]].

\bibitem{Davier:2017zfy}
M.~Davier, A.~Hoecker, B.~Malaescu and Z.~Zhang,
Eur. Phys. J. C \textbf{77}, 827 (2017)
[arXiv:1706.09436 [hep-ph]].

\bibitem{Davier:2010nc}
M.~Davier, A.~Hoecker, B.~Malaescu and Z.~Zhang,
Eur. Phys. J. C \textbf{71}, 1515 (2011)
[erratum: Eur. Phys. J. C \textbf{72}, 1874 (2012)]
[arXiv:1010.4180 [hep-ph]].

\bibitem{Keshavarzi:2018mgv}
A.~Keshavarzi, D.~Nomura and T.~Teubner,
Phys. Rev. D \textbf{97}, 114025 (2018)
[arXiv:1802.02995 [hep-ph]].

\bibitem{Colangelo:2018mtw}
G.~Colangelo, M.~Hoferichter and P.~Stoffer,
JHEP \textbf{02}, 006 (2019)
[arXiv:1810.00007 [hep-ph]].

\bibitem{Hoferichter:2019mqg}
M.~Hoferichter, B.L.~Hoid and B.~Kubis,
JHEP \textbf{08}, 137 (2019)
[arXiv:1907.01556 [hep-ph]].

\bibitem{Davier:2019can}
M.~Davier, A.~Hoecker, B.~Malaescu and Z.~Zhang,
Eur. Phys. J. C \textbf{80}, no.3, 241 (2020)
[erratum: Eur. Phys. J. C \textbf{80}, no.5, 410 (2020)]
[arXiv:1908.00921 [hep-ph]].

\bibitem{Keshavarzi:2019abf}
A.~Keshavarzi, D.~Nomura and T.~Teubner,
Phys. Rev. D \textbf{101}, 014029 (2020)
[arXiv:1911.00367 [hep-ph]].

\bibitem{Kurz:2014wya}
A.~Kurz, T.~Liu, P.~Marquard and M.~Steinhauser,
Phys. Lett. B \textbf{734}, 144 (2014)
[arXiv:1403.6400 [hep-ph]].

\bibitem{Melnikov:2003xd}
K.~Melnikov and A.~Vainshtein,
Phys. Rev. D \textbf{70}, 113006 (2004)
[arXiv:hep-ph/0312226 [hep-ph]].

\bibitem{Masjuan:2017tvw}
P.~Masjuan and P.~Sanchez-Puertas,
Phys. Rev. D \textbf{95}, 054026 (2017)
[arXiv:1701.05829 [hep-ph]].

\bibitem{Colangelo:2017fiz}
G.~Colangelo, M.~Hoferichter, M.~Procura and P.~Stoffer,
JHEP \textbf{04}, 161 (2017)
[arXiv:1702.07347 [hep-ph]].

\bibitem{Hoferichter:2018kwz}
M.~Hoferichter, B.~L.~Hoid, B.~Kubis, S.~Leupold and S.~P.~Schneider,
JHEP \textbf{10}, 141 (2018)
[arXiv:1808.04823 [hep-ph]].

\bibitem{Gerardin:2019vio}
A.~G\'erardin, H.B.~Meyer and A.~Nyffeler,
Phys. Rev. D \textbf{100}, 034520 (2019)
[arXiv:1903.09471 [hep-lat]].

\bibitem{Bijnens:2019ghy}
J.~Bijnens, N.~Hermansson-Truedsson and A.~Rodr\'\i{}guez-S\'anchez,
Phys. Lett. B \textbf{798}, 134994 (2019)
[arXiv:1908.03331 [hep-ph]].

\bibitem{Colangelo:2019uex}
G.~Colangelo, F.~Hagelstein, M.~Hoferichter, L.~Laub and P.~Stoffer,
JHEP \textbf{03}, 101 (2020)
[arXiv:1910.13432 [hep-ph]].

\bibitem{Blum:2019ugy}
T.~Blum, N.~Christ, M.~Hayakawa, T.~Izubuchi, L.~Jin, C.~Jung and C.~Lehner,
Phys. Rev. Lett. \textbf{124}, 132002 (2020)
[arXiv:1911.08123 [hep-lat]].

\bibitem{Colangelo:2014qya}
G.~Colangelo, M.~Hoferichter, A.~Nyffeler, M.~Passera and P.~Stoffer,
Phys. Lett. B \textbf{735}, 90 (2014)
[arXiv:1403.7512 [hep-ph]].
%

\bibitem{Borsanyi:2020mff}
S.~Borsanyi  \textit{et al.}
Nature \textbf{593}, no.7857, 51-55 (2021)
[arXiv:2002.12347 [hep-lat]].
%
\bibitem{Ce:2022kxy}
M.C\`e \textit{et al.}
Phys. Rev. D \textbf{106}, 114502 (2022)
[arXiv:2206.06582 [hep-lat]].
%
\bibitem{Alexandrou:2022amy}
C.~Alexandrou \textit{et al.}
[arXiv:2206.15084 [hep-lat]].
%
\bibitem{Blum:2023qou}
T.~Blum \textit{et al.}
[arXiv:2301.08696 [hep-lat]].
%
\bibitem{Bazavov:2023has}
A.~Bazavov \textit{et al.}
[arXiv:2301.08274 [hep-lat]].
%
\bibitem{Crivellin:2020zul}
A.~Crivellin, M.~Hoferichter, C.A.~Manzari and M.~Montull,
Phys. Rev. Lett. \textbf{125}, 091801 (2020)
[arXiv:2003.04886 [hep-ph]].
%
\bibitem{Keshavarzi:2020bfy}
A.~Keshavarzi, W.J.~Marciano, M.~Passera and A.~Sirlin,
Phys. Rev. D \textbf{102}, 033002 (2020)
[arXiv:2006.12666 [hep-ph]].
%
\bibitem{Colangelo:2020lcg}
G.~Colangelo, M.~Hoferichter and P.~Stoffer,
Phys. Lett. B \textbf{814}, 136073 (2021)
[arXiv:2010.07943 [hep-ph]].



\bibitem{CMS:2019mij}
A.M.~Sirunyan \textit{et al.} [CMS],
Phys. Lett. B \textbf{798}, 134992 (2019)
[arXiv:1907.03152 [hep-ex]].

\bibitem{ATLAS:2019odt}
M.~Aaboud \textit{et al.} [ATLAS],
JHEP \textbf{07}, 117 (2019)
[arXiv:1901.08144 [hep-ex]].

\bibitem{Assamagan:2002kf}
K.A.~Assamagan, A.~Deandrea and P.A.~Delsart,
Phys. Rev. D \textbf{67}, 035001 (2003)
[arXiv:hep-ph/0207302 [hep-ph]].

\bibitem{Omura:2015xcg}
Y.~Omura, E.~Senaha and K.~Tobe,
Phys. Rev. D \textbf{94}, 055019 (2016)
[arXiv:1511.08880 [hep-ph]].

\bibitem{CMS:2019pex}
A.M.~Sirunyan \textit{et al.} [CMS],
JHEP \textbf{03}, 103 (2020)
[arXiv:1911.10267 [hep-ex]].

\bibitem{Crivellin:2017upt}
A.~Crivellin, J.~Heeck and D.~M\"uller,
Phys. Rev. D \textbf{97}, 035008 (2018)
[arXiv:1710.04663 [hep-ph]].

\bibitem{Hou:2019uxa}
W.-S.~Hou, M.~Kohda, T.~Modak and G.-G.~Wong,
Phys. Lett. B \textbf{800}, 135105 (2020)
[arXiv:1903.03016 [hep-ph]].

\bibitem{Crivellin:2013wna}
A.~Crivellin, A.~Kokulu and C.~Greub,
Phys. Rev. D \textbf{87},  094031 (2013)
[arXiv:1303.5877 [hep-ph]].

\bibitem{Belle:2019iji}
M.T.~Prim \textit{et al.} [Belle],
Phys. Rev. D \textbf{101}, 032007 (2020)
[arXiv:1911.03186 [hep-ex]].
%

\bibitem{Iguro:2019sly}
S.~Iguro, Y.~Omura and M.~Takeuchi,
JHEP \textbf{11}, 130 (2019)
[arXiv:1907.09845 [hep-ph]].

\bibitem{Grinstein:1988me}
B.~Grinstein, M.J.~Savage and M.~B.~Wise,
Nucl. Phys. B \textbf{319}, 271 (1989).

\bibitem{Buchalla:1995vs}
G.~Buchalla, A.J.~Buras and M.E.~Lautenbacher,
Rev. Mod. Phys. \textbf{68}, 1125 (1996)
[arXiv:hep-ph/9512380 [hep-ph]].

\bibitem{Chetyrkin:1996vx}
K.G.~Chetyrkin, M.~Misiak and M.~Munz,
Phys. Lett. B \textbf{400}, 206 (1997)
[erratum: Phys. Lett. B \textbf{425}, 414 (1998)]
[arXiv:hep-ph/9612313 [hep-ph]].

\bibitem{Becirevic:2016zri}
D.~Be\v{c}irevi\'c, O.~Sumensari and R.~Zukanovich Funchal,
Eur. Phys. J. C \textbf{76}, 134 (2016)
[arXiv:1602.00881 [hep-ph]].

\bibitem{Gratrex:2015hna}
J.~Gratrex, M.~Hopfer and R.~Zwicky,
Phys. Rev. D \textbf{93}, 054008 (2016)
[arXiv:1506.03970 [hep-ph]].
\bibitem{Ciuchini:1997xe}
M.~Ciuchini, G.~Degrassi, P.~Gambino and G.F.~Giudice,
Nucl. Phys. B \textbf{527}, 21-43 (1998)
[arXiv:hep-ph/9710335 [hep-ph]].
%
\bibitem{Hou:1987kf}
W.-S.~Hou and R.S.~Willey,
Phys. Lett. B \textbf{202}, 591-595 (1988).


\bibitem{Iguro:2017ysu}
S.~Iguro and K.~Tobe,
Nucl. Phys. B \textbf{925} (2017), 560
[arXiv:1708.06176 [hep-ph]].


\end{thebibliography}
\end{document}